%% file: arxiv_Imbach_Pan.tex
\documentclass[runningheads]{llncs}

\usepackage{amssymb,amsmath}
\usepackage{fancyvrb}
\usepackage{color}
\usepackage{bbold}
\usepackage{graphicx}
\usepackage{enumerate}
\newcommand{\RI}[1]{{\color{magenta}#1}}
\renewcommand{\RI}[1]{#1}

\newcommand{\RO}[1]{{\color{blue}#1}}
\renewcommand{\RO}[1]{#1}

\newcommand {\R}   {\mathbb R}

\newcommand {\C}   {\mathbb C} 

\newcommand {\N}   {\mathbb N}

\newcommand{\intbox}{%
	  {\,\setlength{\unitlength}{.33mm}\framebox(4,7){}\,}}
\newcommand {\IR}   {\intbox{\R}}
\newcommand {\IC}   {\intbox{\C}}
\newcommand{\radius}[1]{\textnormal{r}\left(#1\right)}
\renewcommand{\middle}[1]{\textnormal{c}\left(#1\right)}
\newcommand{\width}[1]{\textnormal{w}\left(#1\right)}

\newcommand {\ii}  {{\mathbf i}}
\renewcommand {\Re}  {{\mathcal Re}}
\renewcommand {\Im}  {{\mathcal Im}}
	
\newcommand{\disc}[1]{D(#1)}
\newcommand{\annu}[1]{A\left(#1\right)}
\newcommand{\contDisc}[1]{\disc{#1}}
	
\newcommand{\lcf}[1]{\textnormal{lcf}\left(#1\right)}
\newcommand{\pol}{p}

\newcommand{\roo}[1]{\alpha_{#1}}
\renewcommand{\root}{\alpha}
\newcommand{\var}{z}
\renewcommand{\deg}{d}
\newcommand{\solsIn}[2]{Z\left(#1,#2\right)}
\newcommand{\nbSolsIn}[2]{\#\left(#1,#2\right)}

\newcommand{\logm}[1]{\textnormal{l}\overline{\textnormal{og}}#1}

\newcommand{\cost}[1]{C\left(#1\right)}
\renewcommand{\prec}[1]{L\left(#1\right)}

\newcommand{\calO}{{\mathcal{O}}}
\newcommand{\calI}{{\mathcal{I}}}
\newcommand{\calP}{{\mathcal{P}}}
\newcommand{\OrNumSet}{\calO_{\C}}
\newcommand{\OrNum}[1]{\calO_{#1}}
\newcommand{\OrNumEv}[2]{\calO_{#1}\left({#2}\right)}
\newcommand{\OrPol}[1]{\calI_{#1}}
\newcommand{\OrPolEv}[3]{\calI_{#1}\left({#2},{#3}\right)}
\newcommand{\OrP}{\calP}
\newcommand{\OrPEv}[2]{\calP\left({#1},{#2}\right)}
\newcommand{\OrPp}{\calP'}
\newcommand{\OrPpEv}[2]{\calP'\left({#1},{#2}\right)}
\newcommand{\app}[1]{\widetilde{#1}}

\newcommand{\wt}[1]{\widetilde{#1}}
\newcommand{\PowerSum}[4]{s_{#1}\left({#2},{#3},{#4}\right)}
\newcommand{\PowerSumUnit}[1]{s_{#1}}
\newcommand{\CauchySum}[5]{\wt{s_{#1}}^{#2}\left({#3},{#4},{#5}\right)}
\newcommand{\CauchySumApproxL}[6]{\intbox{\wt{s_{#1}}^{#2}\left({#3},{#4},{#5},{#6}\right)}}

\newcommand{\CauchySumUnit}[2]{\wt{s_{#1}}^{#2}}
\newcommand{\thetaval}{\frac{4}{3}}

\usepackage[noend]{algpseudocode}
\usepackage{algorithm}
\newcommand{\ass}{\leftarrow}

\newcommand{\AppPhs}{{\bf ApproxShs}}
\newcommand{\CauchyRootCounter}{{\bf CauchyRC1}}
\newcommand{\CauchyExclusion}{{\bf CauchyET}}
	
\newcommand{\CauchyRootCounterVerifHeu}{{\bf CauchyRC2}}

\newcommand{\RootRadius}{{\bf RootRadius}}
\newcommand{\Compression}{{\bf Compression}}
\newcommand{\Cauchy}{{\bf CauchyRootFinder}}

	\newcommand{\Mig}[2]{\mbox{\tt Mig}_{#1,#2}}
	\newcommand{\Man}[1]{\mbox{\tt Man}_{#1}}
	\newcommand{\Run}[1]{\mbox{\tt Run}_{#1}}
	
\usepackage{xspace}
\newcommand{\mpsolve}{\texttt{MPsolve}\xspace}
\newcommand{\ccluster}{\texttt{Ccluster}\xspace}
\newcommand{\cclusterAddr}{\url{https://github.com/rimbach/Ccluster}}
\newcommand{\cauchyN}{\texttt{CauchyQIR}\xspace}
\newcommand{\cauchyC}{\texttt{CauchyComp}\xspace}

\newcommand{\clang}{\texttt{C}\xspace}

\newcommand{\machine}{\texttt{Intel(R) Core(TM) i7-8700 CPU @ 3.20GHz}\xspace}

\newcommand{\cored}[1]{{\color{red}#1}}
\newcommand{\coblue}[1]{{\color{blue}#1}}
\newcommand{\forfinal}[1]{{\color{red}#1}}
\renewcommand{\forfinal}[1]{{#1}}

\newtheorem{Definition}{Definition}
\newtheorem{Remark}[Definition]{Remark}
\newtheorem{Theorem}[Definition]{Theorem}
\newtheorem{Corollary}[Definition]{Corollary}
\newtheorem{Proposition}[Definition]{Proposition}
\newtheorem{Lemma}[Definition]{Lemma}

\graphicspath{{./figures}} 

\usepackage{lineno}

\begin{document}

\titlerunning{Accelerated Subdivision Algorithms for Oracle Polynomials}
\title{
Accelerated Subdivision for Clustering Roots of Polynomials given by Evaluation Oracles
}

\authorrunning{R. Imbach and V. Pan}
\author{R\'emi Imbach\inst{1}
  \and Victor Y. Pan\inst{2}
     \institute{Universit\'e de Lorraine, CNRS, Inria, LORIA, F-54000 Nancy, France\\
                \email{remi.imbach@laposte.net},
                \and 
                Lehman College and Graduate Center of City University of New York\\
                \email{victor.pan@lehman.cuny.edu} }
                }
\maketitle


\begin{abstract}
In our quest for the design, the analysis 
and the implementation of a subdivision algorithm
for finding the complex roots of univariate polynomials 
given by oracles for their evaluation, we present 
sub-algorithms allowing 
substantial acceleration of subdivision for complex roots 
clustering for such polynomials.
We rely on Cauchy sums which approximate
power sums of the roots in a fixed complex disc and 
can be computed in a small number
of evaluations --polylogarithmic in the degree.
We describe root exclusion, root counting, 
root radius approximation and 
a procedure for contracting a disc towards the cluster of root it 
contains, called $\varepsilon$-compression.
To demonstrate the efficiency of our algorithms, we combine
them in a prototype root clustering algorithm.
For computing clusters of roots of polynomials 
that can be evaluated fast,
our implementation competes advantageously with 
user's choice for root finding, {\tt MPsolve}.


\end{abstract}

 \section{Introduction}
 \label{sec_introduction}
 
 
 We consider the 
 \begin{center} \fbox{
		\begin{minipage}{0.95 \textwidth} \noindent
		\textbf{$\varepsilon$-Complex Root Clustering Problem ($\varepsilon$-CRC)}\\ \noindent
		\textbf{Given:} a polynomial $\pol\in\C[\var]$ of degree $d$,
		                $\varepsilon>0$\\
		\noindent \textbf{Output:} $\ell\leq d$ couples
		$(\Delta^1,m^1),\ldots,(\Delta^\ell,m^\ell)$ satisfying:\\
		\noindent \hphantom{\textbf{Output:} }- the $\Delta^j$'s are pairwise disjoint discs of radii $\leq \varepsilon$,\\
		\noindent \hphantom{\textbf{Output:} }- for any $1\leq j\leq \ell$, $\Delta^j$ and $3\Delta^j$ contain $m^j>0$ roots of $\pol$, \\
		\noindent \hphantom{\textbf{Output:} }- each complex root of $\pol$ is in a $\Delta^j$ for some $j$.
	\end{minipage}
	} \end{center}
 Here and hereafter 
 \emph{root(s)} stands for
 \emph{root(s) of $\pol$} and are counted with multiplicities,
 $3\Delta^j$ for the factor $3$ concentric dilation of $\Delta^j$,
 and $\pol$ is a {\em Black box polynomial}: its coefficients are not known, but we are given {\em evaluation oracles}, that is, procedures for the evaluation of $\pol$, its derivative $\pol'$
 and hence the ratio $\pol'/\pol$ at a point
 $c\in\C$ with a fixed precision.
 Such a black box polynomial can come from an experimental process
 or can be defined by a procedure, for example Mandelbrot's polynomials, defined inductively as
\[
 \Man{1}(\var) = \var,~~
 \Man{k}(\var) = \var\Man{k-1}(\var)^2 + 1.
\]
$\Man{k}(\var)$ has degree $\deg=2^{k}-1$ and $\deg$ non-zero coefficients 
but can be evaluated fast, i.e. in $O(k)$ arithmetic operations.
Any polynomial given by its coefficients can be 
%
%
 handled as a black box polynomial, 
 and the evaluation subroutines for $\pol, \pol'$ and $\pol'/\pol$ are fast if $\pol$ is sparse or Mandelbrot-like.
 One can solve root-finding problems and in particular the $\varepsilon$-CRC problem
 for black box polynomials
 by first retrieving the coefficients by means of 
evaluation-interpolation, e.g., with FFT 
and inverse FFT, and then by applying the algorithms of
\cite{becker2016complexity,bini2014solving,moroz2021fast,pan2002univariate,sagraloff2016computing}.
Evaluation-interpolation, however,
%
%
 decompresses the representation of a polynomial, which can blow up its input length, in particular, can  destroy sparsity. 
 We do not require knowledge of the coefficients of an input polynomial, but instead use evaluation oracles.
 
 Functional root-finding iterations
 such as Newton's, Weierstrass's (aka Durand-Kerner's) and
 Ehrlich's iterations -- implemented in \mpsolve 
 \cite{bini2014solving} --
 can be applied to 
 approximate the roots of black box polynomials.
 Applying such iterations, however, requires initial points,
 which the known algorithms and in particular \mpsolve 
 obtain by computing root radii, and for that 
 it needs the coefficients of the input polynomial.
 
 \vspace{-0.4cm}
 \subsubsection*{Subdivision algorithms}
 
 Let $\ii$ stand for $\sqrt{-1}$, $c\in\C$, $c=a+\ii b$
 and $r,w\in\R$, $r$ and $w$ positive.
 We call \emph{box} a square complex interval 
 of the form $B(c,w):=[a-\frac{w}{2},a+\frac{w}{2}]+\ii[b-\frac{w}{2},b+\frac{w}{2}]$
 and \emph{disc} $\disc{c, r}$
 the set $\{ x\in\C ~|~ |x-c|\leq r\}$.
 The \emph{containing disc} $\contDisc{B(c,w)}$ of a box $B(c,w)$ is $\disc{c,(3/4)w}$.
 For a $\delta>0$ and a box or a disc $S$, $\delta S$ denotes factor $\delta$ concentric dilation of $S$.
 
 We consider algorithms based on iterative subdivision of 
 an initial box $B_0$ 
 (see \cite{becker2016complexity,becker2018near,pan2000approximating})
 and adopt the framework of \cite{becker2016complexity,becker2018near}
 which relies on two basic subroutines:
 an \emph{Exclusion Test} (ET) -- deciding that a small inflation of a disc contains no root -- 
 and a \emph{Root Counter} (RC) -- counting the number of roots in a small inflation of a disc.
 A box $B$ of the subdivision tree
 is tested for root exclusion or inclusion
 by applying the ET and RC to $\contDisc{B}$,
 which can fail and return $-1$ when $\contDisc{B}$ has 
 some roots near its boundary circle.
 In \cite{becker2016complexity}, ET and RC are based on the Pellet's theorem, 
 requiring the knowlege of the coefficients of $\pol$ and
 shifting the center of considered disc into the origin ({\it Taylor's shifts}); then
 Dandelin-Lobachevsky-Gr\"affe iterations, aka 
 \emph{root-squaring} iterations, enable the 
 following properties for boxes $B$ and discs $\Delta$:
 
 (p1) if $2B$ contains no root, ET applied to $\contDisc{B}$
  returns 0,
  
 (p2) if $\Delta$ and $4\Delta$ contain $m$ roots, RC
  applied to $2\Delta$ returns $m$.
 
 \noindent
 (p1) and (p2) bound the depth of the subdivision tree.
 To achieve quadratic convergence to clusters of roots,
 \cite{becker2016complexity} uses 
 a complex version of the Quadratic Interval Refinement
 iterations of J. Abbott \cite{abbott2014quadratic}, aka
 QIR Abbott iterations, 
 described in details in Algo. 7 of \cite{becker2018near}
 and, like \cite{pan2000approximating}, based on extension 
 of Newton's iterations to multiple roots due to
 Schr\"oder.
 \cite{IPY2018} presents an implementation of 
 \cite{becker2016complexity} in the
 \clang library
 \ccluster\footnote{\cclusterAddr}, which slightly
 outperforms \mpsolve for initial boxes containing 
 only few roots.
 
 In \cite{imbach2020new} we applied 
 an ET based on Cauchy sums approximation.
 It satisfies (p1) and instead of coefficients of $\pol$
 involves 
 $O(\log^2\deg)$ evaluations of $\pol'/\pol$
 with precision $O(d)$ for a disc with radius in $O(1)$;
 although 
%
%
 the output of this ET is only certified if no roots
 lie on or
 near the boundary of the input discs,
 in our extensive experiments it was 
 correct when we dropped this condition. 
 
 \subsection{Our contributions}
 \label{subsec_contributions}
 
 The ultimate goal of our work is to design an algorithm
 for solving the $\varepsilon$-CRC problem for black box polynomials
 which would run faster in practice than the known solvers,
 have low and possibly near optimal Boolean complexity.
 We do not achieve this yet in this paper but rather account
 for the advances along this path by presenting 
 several sub-routines for root clustering.
 We implemented and assembled them in an experimental
 $\varepsilon$-CRC algorithm which outperforms the user's choice software
 for complex root finding, \mpsolve,
 for input polynomials that can be evaluated fast.
 
 \vspace{-0.4cm}
 \subsubsection*{Cauchy ET and RC}
 We describe and analyze a new RC based on Cauchy sum computations
 and satisfying property (p2) which only require the knowledge of 
 evaluation oracles.
 For input disc of radius in $O(1)$, it requires evaluation of $\pol'/\pol$ at $O(\log^2 \deg)$ points with precision $O(d)$ and is based on our ET presented 
 in \cite{imbach2020new}; the support for its correctness is only heuristic.

 \vspace{-0.4cm}
 \subsubsection*{Disc compression}
 For a set $S$, let us write $\solsIn{S}{\pol}$ for the 
 set of roots in $S$
 and $\nbSolsIn{S}{\pol}$ for the cardinality of $\solsIn{S}{\pol}$; 
 two discs $\Delta$ and $\Delta'$ are said
 \emph{equivalent} if $\solsIn{\Delta}{\pol}=\solsIn{\Delta'}{\pol}$.
 We introduce a new sub-problem of $\varepsilon$-CRC:
 \begin{center} \fbox{
		\begin{minipage}{0.95 \textwidth} \noindent
		\textbf{$\varepsilon$-Compression into Rigid Disc 
		($\varepsilon$-CRD)}\\ \noindent
		\textbf{Given:} a polynomial $\pol\in\C[\var]$ of degree $d$,
		                $\varepsilon>0$, $0< \gamma < 1$,\\
		\hphantom{\textbf{Given:} }a disc $\Delta$
		s.t. $\solsIn{\Delta}{\pol}\neq\emptyset$
		and $4\Delta$ is equivalent to $\Delta$.\\
		\noindent \textbf{Output:} a disc $\Delta'\subseteq \Delta$ 
		of radius $r'$ s.t.
		$\Delta'$ is equivalent to $\Delta$ and:\\
		\noindent \hphantom{\textbf{Output:} }- either $r'\leq \varepsilon$,\\
		\noindent \hphantom{\textbf{Output:} }- or $\nbSolsIn{\Delta}{\pol}\geq2$ and $\Delta'$ is at least $\gamma$-\emph{rigid},
		 that is
		 \[
		  \max_{\alpha, \alpha' \in\solsIn{\Delta'}{\pol} }
		  \frac{|\alpha - \alpha'|}{2r'} \geq \gamma.
		 \] 
%
	\end{minipage}
	} \end{center}
 The $\varepsilon$-CRD problem can be solved
 with subdivision and QIR Abbott iteration,
 but this may require, for an initial disk of radius $r$,
 up to $O(\log(r/\max(\varepsilon',\varepsilon))$ calls to the ET in the subdivision
 if the radius of convergence of the cluster in $\Delta$ for Schr\"oder's iteration is in $O(\varepsilon')$.
 
 We present and analyze an algorithm solving the $\varepsilon$-CRD problem
 for $\gamma=1/8$
 based on Cauchy sums approximation and on an algorithm solving the following 
 root radius problem:
 for a given $c\in\C$, a given non-negative integer $m\leq\deg$ and a 
 $\nu>1$,
 find $r$ such that 
 $r_{m}(c,\pol)\leq r \leq \nu r_{m}(c,\pol)$
 where $r_{m}(c,\pol)$ is the smallest radius of a disc centered
 in $c$ and containing exactly $m$ roots of $\pol$.
 Our compression algorithm
 requires only $O(\log\log(r/\varepsilon))$ calls to our RC,
 but a number of evaluations and arithmetic operations
 increasing linearly with $\log (1/\varepsilon)$.
 
 
 \vspace{-0.4cm}
 \subsubsection*{Experimental results}
 
 \begin{table}[t!]
\centering
 \begin{tabular}{cc || c | c | c || c | c | c || c ||}
    &        & \multicolumn{3}{c||}{\cauchyN}  
             & \multicolumn{3}{c||}{\cauchyC} 
             & \mpsolve \\\hline
$\deg$ & $\log_{10}(\varepsilon^{-1})$ 
             & $t$  & $n$  & $t_N$  
             & $t$  & $n$  & $t_C$ 
             & $t$\\\hline 
\multicolumn{9}{c}{Mignotte polynomials, $a=16$} \\\hline
1024 & 5 & \cored{1.68} & 30850 & 0.44 & \coblue{0.96} & 16106 & 0.27 & \cored{1.04} \\
1024 & 10 & \cored{2.08} & 30850 & 0.58 & \coblue{1.07} & 16106 & 0.37 & \cored{1.30} \\
1024 & 50 & \coblue{2.17} & 30850 & 0.71 & \cored{2.70} & 16105 & 1.96 & \cored{4.84} \\
2048 & 5 & \cored{3.84} & 62220 & 0.90 & \coblue{2.13} & 32148 & 0.51 & \cored{4.08} \\
2048 & 10 & \cored{4.02} & 62220 & 1.03 & \coblue{2.36} & 32148 & 0.70 & \cored{5.09} \\
2048 & 50 & \coblue{4.51} & 62220 & 1.25 & \cored{5.62} & 32147 & 3.78 & \cored{17.1} \\
\hline
\multicolumn{9}{c}{Mandelbrot polynomials} \\\hline
1023 & 5 & \cored{10.4} & 30877 & 0.86 & \coblue{6.23} & 18701 & 0.41 & \cored{27.2} \\
1023 & 10 & \cored{10.1} & 30920 & 0.91 & \coblue{6.45} & 18750 & 0.59 & \cored{30.0} \\
1023 & 50 & \cored{10.3} & 30920 & 1.06 & \coblue{8.64} & 18713 & 2.71 & \cored{45.7} \\
2047 & 5 & \cored{24.3} & 62511 & 1.95 & \coblue{15.2} & 39296 & 1.39 & \cored{229.} \\
2047 & 10 & \cored{26.4} & 62952 & 2.31 & \coblue{15.5} & 39358 & 1.71 & \cored{246.} \\
2047 & 50 & \cored{26.1} & 62952 & 2.64 & \coblue{20.4} & 39255 & 6.22 & \cored{380.} \\
\hline
 \end{tabular}
 \caption{Runs of \cauchyN, \cauchyC and \mpsolve on Mignotte and Mandelbrot polynomials. 
 }
\label{table_timings_intro}
\vspace*{-1cm}
\end{table}

 We implemented our algorithms\forfinal{\footnote{\forfinal{they are not publicly realeased yet}}} within \ccluster
 and assembled them in two algorithms named 
 \cauchyN and \cauchyC for solving the
 $\varepsilon$-CRC problem for black box polynomials.
 Both implement the subdivision process of \cite{becker2016complexity}
 with our heuristically correct ET and RC, and
 \begin{itemize}
  \item \cauchyN uses QIR Abbott iterations of \cite{becker2018near} 
  (with Pellet's test replaced by our RC)
  \item \cauchyC uses our compression algorithm instead 
  of QIR Abbott iterations.
 \end{itemize}
 We compare runs of \cauchyN and \cauchyC
 to emphasize the practical improvements allowed by 
 using compression in subdivision algorithms
 for root finding.
 We also compare running times of \cauchyC 
 and \mpsolve to demonstrate that subdivision root finding
 can outperform solvers based on functional
 iterations for polynomials that can be evaluated fast.
 \mpsolve does not cluster roots of a polynomial,
 but approximate each root up to a given error $\varepsilon$.
 Below we used the latest version\footnote{{\tt3.2.1} available 
 here: \url{https://numpi.dm.unipi.it/software/mpsolve}}
 of \mpsolve and call it with:
 {\tt mpsolve -as -Ga -j1 -oN}
 where {\tt N} stands for $\max(1,\lceil\log_{10}(1/\varepsilon)\rceil)$.
 
 All the timings given below have to be understood as sequential
 running times on a \machine machine with Linux.
 We present in table~\ref{table_timings_intro} results obtained
 for Mandelbrot and Mignotte polynomials
 of increasing degree $\deg$ for decreasing error $\varepsilon$.
 The Mignotte polynomial of degree $\deg$
and parameter $a$ is defined as
\[\Mig{\deg}{a}(z)=z^d - 2(2^{\frac{a}{2}-1}z-1)^2.\]
In table~\ref{table_timings_intro}, we account
for the running time $t$ for the three above-mentionned 
solvers. For \cauchyN (resp. \cauchyC),
we also give the number $n$ of exclusion tests in the subdivision
process, and the time $t_N$ (resp. $t_C$)
spent in QIR Abbott iterations (resp. compression).
Mignotte polynomials have two roots with mutual distance close
to the theoretical separation bound; with the $\varepsilon$
used in table~\ref{table_timings_intro}, those roots 
are not separated.

\subsection{Related Work}
 \label{subsec_biblio}
 The subdivision root-finders of  Weyl 1924,  Henrici  1974, Renegar 1987, \cite{becker2018near,pan2000approximating}, rely on ET, RC and root radii sub-algorithms and heavily use the coefficients of $\pol$. 
Design and analysis of subdivision root-finders for a black box  $\pol$ have been continuing since 2018 in \cite{pan2021new} (now over 150 pages), followed by    
 \cite{IP2019,imbach2020new,luan2020faster,pan2019old,pan2020acceleration}
 and this paper. This relies on the novel idea and techniques of compression of a  disc and on novel ET, RC and root radii sub-algorithms. A basic tool of Cauchy sum computation was used in
\cite{schonhage1982fundamental} for polynomial deflation,
but in a large body of our results only Thm. 5 is from \cite{schonhage1982fundamental}; we deduced it in \cite{IP2019,pan2021new} from a new more general theorem of independent interest. Alternative derivation
and analysis of subdivision in \cite{pan2021new}
(yielding a little stronger results but presently not included)
relies on Schr{\"o}der's iterations, extended from \cite{pan2000approximating}. 
The algorithms are analyzed in
\cite{luan2020faster,pan2019old,pan2020acceleration,pan2021new},
 under the model for black box polynomial root-finding of \cite{LV16}.
\cite{IP2019,imbach2020new}
complement this study with some estimates for computational precision and Boolean complexity. We plan to complete them using much more space (cf. 46 pages in each of \cite{schonhage1982fundamental} and \cite{becker2018near}).\footnote{In \cite[Sec. 2]{schonhage1982fundamental},
 called ``The result``, we read: ``The method
is involved and many details still need to be worked out. In this report also
many proofs will be omitted. A full account of the new results shall be given
in a monograph`` which  has actually never appeared. \cite{becker2018near} deduced {\em a posteriori estimates}, depending on  root separation and Mahler’s measure, that is, on the roots themselves, not known a priori.} Meanwhile we borrowed from \cite{becker2018near}
 Pellet's RC (involving  coefficients), Abbott's QIR and the general subdivision algorithm with connected components of boxes extended from \cite{pan2000approximating,renegar1987worst}. 
With our novel sub-algorithms, however,  we significantly 
outperform \mpsolve for polynomials that can be evaluated fast; all previous subdivision root-finders have never come close to such level. 
 \mpsolve relies on Ehrlich's (aka Aberth's) iterations, whose Boolean complexity is proved to be unbounded  because  iterations diverge for  worst case inputs \cite{reinke2020diverging}, but
 divergence never occurs in decades
 of extensive application of these iterations.
 
 \subsection{Structure of the paper}
 \label{subsec_structure}
 In Sec.~\ref{sec_cauchySums}, we describe
 power sums and their approximation with Cauchy sums.
 In Sec.~\ref{sec_exclusion_counting}, we present and analyze
 our Cauchy ET and RC.
 Sec.~\ref{sec_root_radii} is devoted to root radii algorithms
 and Sec.~\ref{sec_compression} to the presentation of 
 our algorithm solving the $\varepsilon$-CRD problem.
 We describe the experimental solvers \cauchyN and \cauchyC
 in Sec.~\ref{sec_root_finders}, numeric results in Sec.~\ref{sec_experiments}
 and conclude in Sec.~\ref{sec_conclusion}.
 We introduce additional definitions and properties in the rest 
 of this section.
 
 \subsection{Definitions and two evaluations bounds}
 \label{subsec_definitions}
 Troughout this paper, $\log$ is the binary logarithm
 and for a positive real number $a$, let $\logm{a}=\max(1,\log a)$.
 
 \vspace{-0.4cm}
  \subsubsection{Annuli, intervals}
 
For $c\in\C$ and positives $r\leq r'\in\R$, 
the annulus $\annu{c,r,r'}$
is the set $\{ \var\in\C ~|~ r'\leq |\var-c|\leq r' \}$.

Let $\IR$ be the set $\{ [a-\frac{w}{2}, a+\frac{w}{2}] ~|~ a,w\in\R, w\geq 0 \}$ of real intervals. For 
$\intbox{a}=[a-\frac{w}{2}, a+\frac{w}{2}]\in\IR$
the center $\middle{\intbox{a}}$,
the width $\width{\intbox{a}}$ 
and the radius $\radius{\intbox{a}}$
of $\intbox{a}$ are respectively
$a$, $w$ and $w/2$.

Let $\IC$ be the set 
$\{ \intbox{a} + \ii\intbox{b} | \intbox{a},\intbox{b}\in \IR \}$ of complex intervals.
If $\intbox{c}\in\IC$, then $\width{\intbox{c}}$
(resp. $\radius{\intbox{c}}$)
is $\max( \width{\intbox{a}}, \width{\intbox{b}} )$
(resp $\width{\intbox{c}}/2$).
The center $\middle{\intbox{c}}$ of $\intbox{c}$ is 
$\middle{\intbox{a}} + \ii\middle{\intbox{b}}$.
 
%
%

%

\vspace{-0.4cm}
\subsubsection{Isolation and rigidity of a disc}
are defined as follows \cite{pan2000approximating,pan2021new}.

\begin{Definition}[Isolation]
 \label{def_isolation_ratio}
 Let $\theta>1$.
 The disc $\Delta=\disc{c,r}$ has {\em isolation} 
 $\theta$ for 
 a polynomial $\pol$ or equivalently 
 is {\em at least} $\theta$-{\em isolated}
 if 
 $\solsIn{\frac{1}{\theta}\Delta}{\pol}=\solsIn{\theta\Delta}{\pol}$,
 that is 
 $\solsIn{\annu{c,r/\theta,r\theta}}{\pol}=\emptyset$.
\end{Definition}

\begin{Definition}[Rigidity]
 \label{def:rigidity}
 For a disc $\Delta=D(c,r)$, define
 \[
    \gamma(\Delta)=\max_{\alpha, \alpha'\in\solsIn{\Delta}{\pol}}
		  \frac{|\alpha - \alpha'|}{2r}
  \]
 and remark that $\gamma(\Delta)\leq 1$.
 We say that $\Delta$ has rigidity $\gamma$
 or equivalently is {\em at least }$\gamma$-{\em rigid}
 if $\gamma(\Delta)\geq \gamma$.
\end{Definition}

\vspace{-0.4cm}
\subsubsection{Oracle Numbers and Oracle Polynomials}
 
 Our algorithms
 deal with numbers that can be approximated arbitrarily closely by a Turing machine.
 We call such approximation automata
 \emph{oracle numbers} and formalize them through
 interval arithmetic. 
 
 For $a\in\C$ 
 we call \emph{oracle} for $a$ a function
 $\OrNum{a}:\N\rightarrow\IC$
 such that 
 $a\in\OrNumEv{a}{L}$ and 
 $\radius{\OrNumEv{a}{L}}\leq2^{-L}$
 for any $L\in\N$.
 In particular, one has
 $| \middle{\OrNumEv{a}{L}} - a |\leq2^{-L}$.
 Let $\OrNumSet$ be the set of oracle numbers
 which can be computed with a Turing machine.
 For a polynomial $\pol\in\C[\var]$, we call
 \emph{evaluation oracle} for $\pol$
 a function
 $\OrPol{\pol}: (\OrNumSet,\N)\rightarrow\IC$,
 such that if $\OrNum{a}$ is an oracle for $a$
 and $L\in\N$, then
 $\pol(a)\in\OrPolEv{\pol}{\OrNum{a}}{L}$
 and 
 $\radius{\OrPolEv{\pol}{\OrNum{a}}{L}}\leq 2^{-L}$.
 In particular, one has
 $| \middle{\OrPolEv{\pol}{\OrNum{a}}{L}} - p(a) |\leq2^{-L}$.
 
 Consider evaluation oracles $\OrPol{\pol}$ and $\OrPol{\pol'}$
 for $\pol$ and $\pol'$.
 If $\pol$ is given by $\deg'\leq\deg+1$ oracles for its coefficients,
 one can easily construct $\OrPol{\pol}$ and $\OrPol{\pol'}$
 by using, for instance, Horner's rule.
 However for procedural polynomials (\emph{e.g.} Mandelbrot), 
 fast evaluation oracles $\OrPol{\pol}$ and $\OrPol{\pol'}$
 are built from procedural definitions.
 
 To simplify notations, we let $\OrPolEv{\pol}{a}{L}$
 stand for $\OrPolEv{\pol}{\OrNum{a}}{L}$.
 In the rest of the paper, 
 $\OrP$ (resp. $\OrPp$) is an evaluation oracle
 for $\pol$ (resp. $\pol'$);
 $\OrPEv{a}{L}$ (resp. $\OrPpEv{a}{L}$) will
 stand for $\OrPolEv{\pol}{\OrNum{a}}{L}$ 
 (resp. $\OrPolEv{\pol'}{\OrNum{a}}{L}$).

%
 
 \vspace{-0.4cm}
 \subsubsection{Two evaluation bounds}
 \label{susubbsec_evaluation_bounds}
 
 The Lemma below \RO{is proved in \ref{app_subsec_proof_eval_bounds} and }
 provides estimates for values of $|\pol|$ and $|\pol'/\pol|$ on the boundary 
 of isolated discs.
 
 \begin{Lemma}
 \label{Lem_eval_bounds}
 Let $\disc{c,r}$ be at least $\theta$-isolated,
 $\var\in\C$, $|\var|=1$ and $g$ be a positive integer. 
 Let $\lcf{\pol}$ be the leading coefficient of $\pol$. Then
 \begin{equation*}
  |\pol(c+r\var^g)| \geq |\lcf{\pol}|\frac{r^\deg(\theta-1)^\deg}{\theta^\deg}
  \text{ ~~and~~ }
  \left|\frac{\pol'(c+r\var^g)}{\pol(c+r\var^g)}\right|
 \leq
 \frac{\deg\theta}{r(\theta-1)}.
 \end{equation*}
%
\end{Lemma}

\section{Power Sums and Cauchy Sums}
 \label{sec_cauchySums}

 \begin{Definition}[Power sums of the roots in a disc]
\label{def_power_sums}
The $h$-th power sum of (the roots of) $\pol$ in the disc $\disc{c,r}$ is the complex number
\begin{equation}
\label{eq_power_sum_definition}
 \PowerSum{h}{\pol}{c}{r}
 =\sum\limits_{\root\in\solsIn{\Delta}{\pol}} \nbSolsIn{\root}{\pol}\root^h
\end{equation}
where $\nbSolsIn{\root}{\pol}$ stands for the multiplicity 
of $\root$ as a root of $\pol$.
\end{Definition}
The power sums $\PowerSum{h}{\pol}{c}{r}$ are equal to Cauchy's integrals 
over the boundary circle $\partial\disc{c,r}$;
by following \cite{schonhage1982fundamental}
they can be approximated by Cauchy sums obtained by
means of the discretization of the integrals:
let $q\geq 1$ be an integer
and $\zeta$ be a primitive $q$-th root of unity.
When $\pol(c+r\zeta^g)\neq 0$ for $g=0,\ldots,q-1$,
and in particular when $\disc{c,r}$ is at least
$\theta$-isolated with $\theta>1$,
define the Cauchy sum 
$\CauchySum{h}{q}{\pol}{c}{r}$
as
\begin{equation}
\label{eq_cauchy_sum_definition}
 \CauchySum{h}{q}{\pol}{c}{r}
 =\frac{r}{q}\sum\limits_{g=0}^{q-1} \zeta^{g(h+1)} 
 \frac{\pol'(c+r\zeta^g)}{\pol(c+r\zeta^g)}.
\end{equation}

For conciseness of notations, we write $\PowerSumUnit{h}$
for 
$\PowerSum{h}{\pol}{0}{1}$
and $\CauchySumUnit{h}{q}$
for 
$\CauchySum{h}{q}{\pol}{0}{1}$.
The following theorem, proved in 
\cite{imbach2020new,schonhage1982fundamental}, 
allows us to approximate power sums by Cauchy sums
in $\disc{0,1}$.

\begin{Theorem}
\label{thm_app_ps_unit_disc}
For  $\theta>1$ and 
integers $h,q$ s.t. $0\leq h < q$ 
 let the unit disc $\disc{0,1}$ be at least
 $\theta$-isolated and contain $m$ roots of $\pol$.
Then
 \begin{equation}
 \label{eq_thm_app_ps_unit_disc}
  |\CauchySumUnit{h}{q}-\PowerSumUnit{h}| \leq \dfrac{m\theta^{-h}+ (\deg-m)\theta^{h}}
                         {\theta^{q}-1}.
 \end{equation}
 \begin{equation}
 \label{eq_thm_app_psh_unit_disc}
  \text{Fix }e>0\text{. If }
  q\geq\lceil \log_{\theta}(\frac{\deg}{e}) \rceil + h +1
        \text{ then }|\CauchySumUnit{h}{q}-\PowerSumUnit{h}| \leq e.
 \end{equation}
\end{Theorem}

Remark that
$\PowerSum{0}{\pol}{c}{r}$ is the number
of roots of $\pol$ in $\disc{c,r}$
and 
$\PowerSum{1}{\pol}{c}{r}/m$ is 
their center of gravity
when $m=\nbSolsIn{\disc{c,r}}{\pol}$.
\medskip

Next we extend Thm.~\ref{thm_app_ps_unit_disc} to the approximation
of $0$-th 
and $1$-st 
power sums by Cauchy sums in any disc,
and define and analyze our basic algorithm 
for the computation of these power sums. 

\subsection{Approximation of the power sums}
\label{subsec_app_power_sums_any_disc}

Let $\Delta=\disc{c,r}$ and define $\pol_\Delta(\var)$ as
$\pol(c+r\var)$ so that
$\root$ is a root of $\pol_\Delta$ in $\disc{0,1}$
if and only if $c+r\root$ is a root of $\pol$ in $\Delta$.
Following Newton's identities, one has:
\begin{align}
\label{eq_pzero_unit_any}
 \PowerSum{0}{\pol}{c}{r} &= \PowerSum{0}{\pol_\Delta}{0}{1},\\
\label{eq_pone_unit_any}
 \PowerSum{1}{\pol}{c}{r} &= c\PowerSum{0}{\pol_\Delta}{0}{1} 
                             + r\PowerSum{1}{\pol_\Delta}{0}{1}.
 \end{align}
Next since $\pol_{\Delta}'(\var)=r\pol'(c+r\var)$,
one has
\[\CauchySum{h}{q}{\pol}{c}{r}
= \frac{1}{q}\sum\limits_{g=0}^{q-1} \zeta^{g(h+1)} 
  \frac{\pol_{\Delta}'(\zeta^g)}{\pol_{\Delta}(\zeta^g)}
=\CauchySum{h}{q}{\pol_{\Delta}}{0}{1}\]
and can easily prove:

\begin{Corollary}[of thm.~\ref{thm_app_ps_unit_disc}]
\label{cor_app_ps_any_disc}
 Let $\Delta=\disc{c,r}$ be at least $\theta$-isolated.
 Let $q>1$, $s_0^* = \CauchySum{0}{q}{\pol}{c}{r}$
 and $s_1^* = \CauchySum{1}{q}{\pol}{c}{r}$.
 Let $e>0$.
 One has
 \begin{equation}
 \label{eq_cor_app_ps_1d0} 
  |s_0^*-\PowerSum{0}{\pol}{c}{r}| 
  \leq \dfrac{\deg}{\theta^{q}-1}.
 \end{equation}
 \begin{equation}
 \label{eq_cor_app_ps_2d0}
  \text{If }
  q\geq\lceil \log_{\theta}(1 + \frac{\deg}{e}) \rceil
        \text{ then }|s_0^*-\PowerSum{0}{\pol}{c}{r}| \leq e.
 \end{equation}
 Let $\Delta$ contain $m$ roots.
 \begin{equation}
 \label{eq_cor_app_ps_1d1}
  |mc+rs_1^*-\PowerSum{1}{\pol}{c}{r}| \leq \dfrac{r\deg\theta}{\theta^{q}-1}.
 \end{equation}
 \begin{equation}
 \label{eq_cor_app_ps_2d1}
  \text{If }
  q\geq\lceil \log_{\theta}(1+\frac{r\theta\deg}{e}) \rceil
        \text{ then }|mc+rs_1^*-\PowerSum{1}{\pol}{c}{r}| \leq e.
 \end{equation}
\end{Corollary}

\subsection{Computation of Cauchy sums}

Next we suppose that $\disc{c,r}$ and $q$
are such that $\pol(c+r\zeta^g)\neq 0$ $\forall 0\leq g < q$,
so that $\CauchySum{h}{q}{\pol}{c}{r}$ is well
defined.
We approximate Cauchy sums with 
evaluation oracles $\OrP$, $\OrPp$ by 
choosing a sufficiently large $L$ and
computing
the complex interval:
\begin{equation}
\label{eq_ball_cauchy_sum}
 \CauchySumApproxL{h}{q}{\pol}{c}{r}{L} =
                \frac{r}{q}\sum\limits_{g=0}^{q-1} 
                \OrNum{\zeta^{g(h+1)}}(L) 
                \frac{\OrPpEv{c+r\zeta^g}{L}}
                {\OrPEv{c+r\zeta^g}{L}}.
\end{equation}
$\CauchySumApproxL{h}{q}{\pol}{c}{r}{L}$
is well defined for
$L > \max_{0\leq g < q} \left(-\log_2 ( \pol(c+r\zeta^g) )\right)$
and contains $\CauchySum{h}{q}{\pol}{c}{r}$.
%
%
%
The following result specifies 
$L$ for which we obtain that
$\radius{\CauchySumApproxL{h}{q}{\pol}{c}{r}{L}}\leq e$
for an $e>0$. 
\RO{It is proved in~\ref{app_subsec_proof_lem_prec_approx_sh}.}
\begin{Lemma}
\label{lem_prec_approx_sh}
For strictly positive integer $d$, reals $r$ and $e$ and
$\theta>1$, let
 \begin{align*}
 \prec{\deg,r,e,\theta} &:= \max \left(
        (d+1) \log{ \frac{\theta}{er(\theta-1)} }
        + \log(26rd),1
        \right)\\
        & \in
    O\left(
          \deg\left(\logm{\frac{1}{re}} + \log\frac{\theta}{\theta-1} \right)
    \right).
 \end{align*}
If $L\geq \prec{\deg,r,e, \theta}$
then $\radius{\CauchySumApproxL{h}{q}{\pol}{c}{r}{L}}
 \leq e$.
\end{Lemma}

In the sequel let $\prec{\deg,r}$ stand for $\prec{\deg,r,1/4,2}$.
%

\subsection{Approximating the power sums 
            $\PowerSumUnit{0},\PowerSumUnit{1}, \ldots, \PowerSumUnit{h}$}
Our Algo.~\ref{algo:approximating_phs} computes, for a given integer $h$, 
approximations to power sums $\PowerSumUnit{0}, \PowerSumUnit{1}, \ldots, \PowerSumUnit{h}$ (of $\pol_\Delta$ in $\disc{0,1}$)
up to an error $e$,
based on eqs.~(\ref{eq_cauchy_sum_definition}) 
and~(\ref{eq_thm_app_psh_unit_disc}).

 \begin{algorithm}[h!]
	\begin{algorithmic}[1]
	\caption{$\AppPhs(\OrP, \OrPp, \Delta, \theta, h, e)$ }
	\label{algo:approximating_phs}
	\Require{$\OrP$, $\OrPp$
	         evaluation oracles for $\pol$ and $\pol'$,
	         s.t. $\pol$ is monic of degree $\deg$.
	         $\Delta=D(c,r)$, 
             $\theta\in\R, \theta>1$,
             $h\in\N$, $h\geq0$,
	         $e\in\R$, $e>0$.
	         } 
	\Ensure{
	a flag $success\in\{true, false\}$, a vector
	$[\intbox{s_0}, \ldots, \intbox{s_h}]$.
    }
	
	\State $e'\ass e/4$, 
	       $q\ass \lceil \log_{\theta}(4d/e) \rceil + h + 1$
	\State $\ell\ass \frac{r^\deg(\theta-1)^\deg}{\theta^d}$,
	       $\ell'\ass \frac{\deg\theta}{r(\theta-1)}$
	\State $L\ass 1$
	\State $[\intbox{s_0}, \ldots, \intbox{s_h}]
	       \ass [\C,\ldots, \C]$
	\While{ $\exists i\in\{0,\ldots, h\}$ s.t. $\width{\intbox{s_i}}\geq e$ }
        \State $L\ass 2L$
	    \For {$g=0, \ldots, q-1$}
            \State Compute intervals 
                $\OrPEv{c+r\zeta^g}{L}$
                and 
                $\OrPpEv{c+r\zeta^g}{L}$
        \EndFor
        \If{$\exists g\in\{0,\ldots, q-1\}$ s.t. $|\OrPEv{c+r\zeta^g}{L}|<\ell$
            {\bf or} $\left|\frac{\OrPpEv{c+r\zeta^g}{L}}
           {\OrPEv{c+r\zeta^g}{L}}\right|>\ell'$}
            \State  \Return $false$, $[\intbox{s_0}, \ldots, \intbox{s_h}]$
        \EndIf
        \If{$\exists g\in\{0,\ldots, q-1\}$ s.t. $\frac{\ell}{2}\in|\OrPEv{c+r\zeta^g}{L}|$ {\bf or}
            $2\ell'\in \left|\frac{\OrPpEv{c+r\zeta^g}{L}}
           {\OrPEv{c+r\zeta^g}{L}}\right|$}
            \State {\bf continue}
        \EndIf
        
        \For {$i=0, \ldots, h$}
        \State 
        $\intbox{s_i^*}\ass\CauchySumApproxL{i}{q}{\pol}{c}{r}{L}$
       \hfill{\it //as in eq.(\ref{eq_ball_cauchy_sum})}
       \State $\intbox{s_i}\ass \intbox{s_i^*} + [-e',e']+\ii [-e',e']$
       \EndFor
	\EndWhile
	\State \Return $true$, $[\intbox{s_0}, \ldots, \intbox{s_h}]$
	\end{algorithmic}
\end{algorithm}
\vspace{-0.3cm}

Algo.~\ref{algo:approximating_phs} satisfies the
following proposition\RO{~proved in~\ref{app_subsec_proof_algo_compute_power_sums}}.

\begin{Proposition}
 \label{prop_algo_compute_power_sums}
 Algo.~\ref{algo:approximating_phs} terminates for an 
 $L\leq \prec{\deg, r, e/4, \theta}$.\\
 Let $\AppPhs(\OrP, \OrPp, \Delta, \theta, h, e)$
 return 
 $(success, [\intbox{s_0}, \ldots, \intbox{s_h}])$.
 Let $\Delta=\disc{c,r}$ and 
 $\pol_{\Delta}(\var) = \pol(c+r\var)$.
 If $\theta>1$,
 one has:
 \vspace{-0.1cm}
 \begin{enumerate}[(a)]
  \item If $A(c,r/\theta, r\theta)$ contains no root of $\pol$,
        then $success = true$ and for all $i\in\{0,\ldots, h\}$, 
        $\width{\intbox{s_i}}<e$ and 
        $\intbox{s_i}$ contains 
        $\PowerSum{i}{\pol_\Delta}{0}{1}$.
        
  \item If $e\leq 1$ and $\disc{c,r\theta}$ contains no root of $\pol$
        then $success = true$ and for all $i\in\{0,\ldots, h\}$, 
        $\intbox{s_i}$ contains the unique integer $0$.
        
  \item If $e\leq 1$ and $A(c,r/\theta, r\theta)$ contains no root of $\pol$,
        $\intbox{s_0}$ contains the unique integer
        $\PowerSum{0}{\pol}{c}{r}=\nbSolsIn{\Delta}{\pol}$.
        
  \item If $success = false$,
        then $A(c,r/\theta, r\theta)$ and $\disc{c,r\theta}$ contain 
        (at least) a root of $\pol$.
        
  \item If $success = true$ and $\exists i\in\{0,\ldots, h\}$, 
        s.t. $\intbox{s_i}$ does not contain $0$
        then $A(c,r/\theta, r\theta)$ and $\disc{c,r\theta}$ contains 
        (at least) a root of $\pol$.
\end{enumerate}
\end{Proposition}

\section{Exclusion Test and Root Counters}
 \label{sec_exclusion_counting}
 
 In this section we define and analyse
 our base tools for
 disc exclusion and root counting.
 We recall in subsec.~\ref{subsection_unsure_counting}
 and subsec.~\ref{subsection_heuristic_exclusion}
 the RC and the ET presented in \cite{imbach2020new}.
 In subsec.~\ref{subsection_heuristic_root_counting},
 we propose a heuristic certification of root counting
 in which the assumed isolation for a disc $\Delta$ is 
 heuristically verified by applying 
 sufficiently many ETs on the contour of $\Delta$.
 
 For $\deg\geq 1, r>0$ and $\theta>1$, define 
 \begin{equation}
 \label{eq_cost}
  \cost{\deg,r,e,\theta} := \log(\prec{\deg,r,e,\theta})\log_\theta(\deg/e)
 \end{equation}
 and $\cost{\deg,r} = \cost{\deg,r,1/4,2}$.
 
 \subsection{Root Counting 
 with known isolation}
 \label{subsection_unsure_counting}
 
 For a disc $\Delta$ which is at least $\theta$-isolated for $\theta>1$,
 algo.~\ref{algo:unsure_counting}
 computes the number $m$
 of roots in $\Delta$
 as the unique integer in the interval of width $<1$ obtained 
 by approximating $0$-th cauchy sum of $\pol_\Delta$
 in the unit disc within error $<1/2$.
 
   \begin{algorithm}[h!]
	\begin{algorithmic}[1]
	\caption{$\CauchyRootCounter(\OrP, \OrPp, \Delta, \theta)$ }
	\label{algo:unsure_counting}
	\Require{$\OrP$, $\OrPp$
	         evaluation oracles for $\pol$ and $\pol'$,
	         s.t. $\pol$ is monic of degree $\deg$.
	         $\Delta=D(c,r)$, 
             $\theta\in\R, \theta>1$.
	         } 
	\Ensure{ An integer $m\in\{ -1, 0, \ldots, \deg \}$.
           }
    
    \State $(success,[\intbox{s_0}])\ass \AppPhs(\OrP, \OrPp, \Delta, \theta, 0, 1)$
    \If{$success=false$ or $\intbox{s_0}$ contains no integer}
        \State \Return $-1$
    \EndIf
    \State \Return the unique integer in $\intbox{s_0}$
	\end{algorithmic}
\end{algorithm}
\vspace{-0.3cm}

\begin{Proposition}
\label{prop_CauchyRootCounter}
 Let $\Delta=\disc{c,r}$.
 $\CauchyRootCounter(\OrP, \OrPp, \Delta, \theta)$
 requires evaluation of~$\OrP$ and $\OrPp$ at
 $O(\cost{\deg,r,1,\theta})$
 points and $O(\cost{\deg,r,1,\theta})$
 arithmetic operations, all 
 with precision less than $\prec{\deg, r,1/4, \theta}$.
 Let $m$ be the output of the latter call.
 \vspace{-0.3cm}
 \begin{enumerate}[(a)]
  \item If $\annu{c,r/\theta, r\theta}$ contains no roots of $\pol$
  then $m=\nbSolsIn{\Delta}{\pol}$.
  \item If $m\neq 0$ then $\pol$ has a root in the disc
  $\theta\Delta$.
 \end{enumerate}

\end{Proposition}

Prop.~\ref{prop_CauchyRootCounter} is a direct consequence 
of Prop.~\ref{prop_algo_compute_power_sums}:
in each execution of the while loop in 
$\AppPhs(\OrP, \OrPp, \Delta, \theta, 0, 1)$,
$\OrP$ and $\OrPp$ are evaluated at $O(\log_{\theta}\deg/e)$
points and the while loop executes an
$O(\log( \prec{\deg, r, 1, \theta} ))$ number of times.

\subsection{Cauchy Exclusion Test}
 \label{subsection_heuristic_exclusion}
 
We follow \cite{imbach2020new}
and increase the chances for obtaining a correct result
for the exclusion of a disc with unknown isolation
by approximating the first three 
power sums of $\pol_\Delta$ in $\disc{0,1}$
in Algo.~\ref{algo:unsure_exclusion}. One has:

\begin{Proposition}
 \label{prop_CauchyExclusion}
 Let $\Delta=\disc{c,r}$.
 $\CauchyExclusion(\OrP, \OrPp, \Delta)$
 requires evaluation of $\OrP$ and $\OrPp$ at
 $O(\cost{\deg, r})$
 points and $O(\cost{\deg, r})$
 arithmetic operations, all 
 with precision less than $\prec{\deg, r}$.
 Let $m$ be the output of the latter call.
 \vspace{-0.3cm}  
 \begin{enumerate}[(a)]
  \item If $\disc{c, 4r/3}$ contains no roots of $\pol$
  then $m=0$. Let $B$ be a box so that $2B$ contains no root
  and suppose $\Delta=\disc{B}$;
  then $m=0$.
  \item If $m\neq 0$ then $\pol$ has a root in the disc
   $(4/3)\Delta$.
 \end{enumerate}
\end{Proposition}

\vspace{-0.3cm}
 \begin{algorithm}[h!]
	\begin{algorithmic}[1]
	\caption{$\CauchyExclusion(\OrP, \OrPp, \Delta)$ }
	\label{algo:unsure_exclusion}
	\Require{$\OrP$, $\OrPp$
	         evaluation oracles for $\pol$ and $\pol'$,
	         s.t. $\pol$ is monic of degree $\deg$.
	         $\Delta=D(c,r)$.
	         } 
	\Ensure{ An integer $m\in\{ -1, 0\}$.
           }
    
    \State $(success,[\intbox{s_0}, \intbox{s_1} ,\intbox{s_2}])\ass \AppPhs(\OrP, \OrPp, \Delta, 4/3, 2, 1)$
    \If{$success=false$ \bf{or}
        $0\notin\intbox{s_0}$ \bf{or} 
          $0\notin\intbox{s_1}$ \bf{or} 
          $0\notin\intbox{s_2}$
    }
        \State \Return $-1$
    \EndIf
    \State \Return $0$
	\end{algorithmic}
\end{algorithm}
\vspace{-0.3cm}

\subsection{Cauchy Root Counter}
 \label{subsection_heuristic_root_counting}
 
\begin{figure}
	 \centering
	  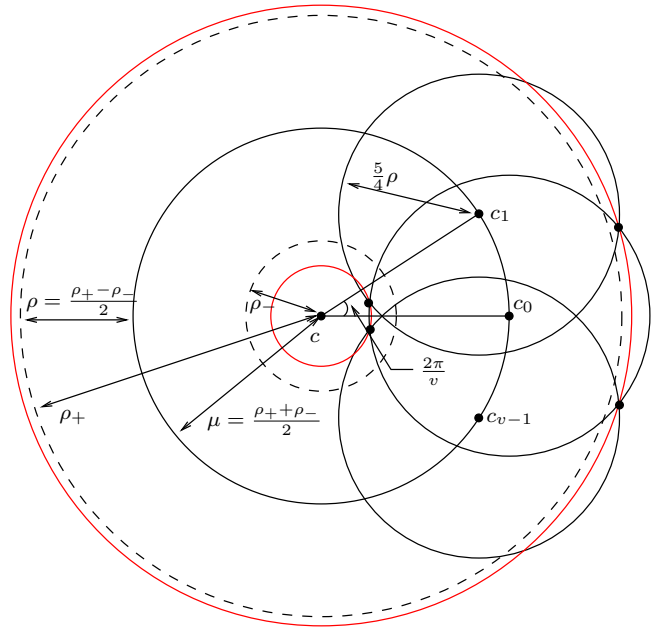
	    \caption{Illustration for Lem.~\ref{lem_circles}.
	    In red, the inner and outer circles of the annulus
	    covered
	    by the $v$ discs $\disc{c_j,(5/4)\rho}$.}
	 \label{fig:countingTest}
	\end{figure}

 We begin with a lemma 
 \RO{proved in \ref{app_subsec_proof_lem_circles} and}
 illustrated 
 in Fig.~\ref{fig:countingTest}.

 \begin{Lemma}
\label{lem_circles}
 Let $c\in\C$ and $\rho_-,\rho_+\in\R$.
 Define $\mu=\frac{\rho_+ + \rho_-}{2}$,
 $\rho=\frac{\rho_+ - \rho_-}{2}$,
 $w=\frac{\mu}{\rho}$,
 $v=\lceil 2\pi w \rceil$
 and $c_j = c + \mu e^{j\frac{2\pi\ii}{v}}$
 for $j=0,\ldots,v-1$.
 Then the re-union of the discs $\disc{c_j,(5/4)\rho}$ covers
 the annulus $\annu{c, \rho_-, \rho_+}$.
\end{Lemma}

For a disc $\disc{c,r}$ and a given $a>1$, we follow 
Lem.~\ref{lem_circles} and cover
the annulus $\annu{c, r/a, ra}$ with $v$ discs of radius
$r\frac{5(a-1/a)}{4*2}$
centered at $v$ equally spaced points of the boundary circle
of
$\disc{c,r\frac{a+1/a}{2}}$.
Define 
 \begin{equation}
 \label{eq:fmoins}
  f_-(a,\theta) = \frac{1}{2}(a(1-\frac{5}{4}\theta) + \frac{1}{a}(1+\frac{5}{4}\theta))
 \end{equation}
 and
 \begin{equation}
 \label{eq:fplus}
  f_+(a,\theta) = \frac{1}{2}(a(1+\frac{5}{4}\theta) + \frac{1}{a}(1-\frac{5}{4}\theta)),
 \end{equation}
 then the annulus $\annu{c, rf_-(a,\theta), rf_+(a,\theta)}$ covers
 the $\theta$-inflation of those $v$ discs.
 
Algo.~\ref{algo:CauchyRootCounterVerifHeu}
counts the number of roots of $\pol$ in 
a disc and satisfies:

\begin{algorithm}[h!]
	\begin{algorithmic}[1]
	\caption{$\CauchyRootCounterVerifHeu(\OrP, \OrPp, \Delta, a)$ }
	\label{algo:CauchyRootCounterVerifHeu}
	\Require{$\OrP$, $\OrPp$
	         evaluation oracles for $\pol$ and $\pol'$,
	         s.t. $\pol$ is monic of degree $\deg$.
	         $\Delta=D(c,r)$.
	         $a\in\R, a>1$.
	         } 
	\Ensure{ An integer $m\in\{ -1, 0, \ldots, \deg \}$. }
    
    \noindent
    \coblue{{\it Verify that $\Delta$ is at least $a$-isolated with \CauchyExclusion}}
	\State $\rho_-\ass \frac{1}{a}r$, $\rho_+ = ar$.
	\State $\rho\ass\frac{\rho_+-\rho_-}{2}$,
	       $\mu\ass\frac{\rho_+ + \rho_-}{2}$,
	       $w\ass\frac{\mu}{\rho}$, $v\ass\lceil 2\pi w\rceil$, $\zeta\ass\exp(\frac{2\pi\ii}{v})$
    \For{ $i=0,\ldots,v-1$ }
        \State $c_i\ass c + \mu\zeta^i$
        \If{ $\CauchyExclusion(\OrP, \OrPp, \disc{c_i,\frac{5}{4}\rho})$ returns $-1$ }
            \State \Return $-1$
            \hfill{\it // $\annu{c,rf_-(a,\thetaval), rf_+(a,\thetaval)}$ contains a root}
        \EndIf
    \EndFor

    \noindent
    \coblue{{\it $\Delta$ is at least $a$-isolated according to \CauchyExclusion}}
	\State \Return $\CauchyRootCounter(\OrP, \OrPp, \Delta, a)$
	\end{algorithmic}
\end{algorithm}
\vspace{-0.3cm}

\begin{Proposition}
 \label{prop_CauchyRootCounterVerifHeu}
 The call $\CauchyRootCounterVerifHeu(\OrP, \OrPp, \Delta, a)$ amounts to 
 $\lceil 2\pi\frac{a^2+1}{a^2-1}\rceil$
 calls to $\CauchyExclusion$
 and one call to $\CauchyRootCounter$.
 
 Let $\Delta=\disc{c,r}$ and $A$ be the annulus
 $\annu{c,rf_-(a,\thetaval), rf_+(a,\thetaval)}$.
 Let $m$ be the output of the latter call.
 \vspace{-0.3cm}
 \begin{enumerate}[(a)]
  \item If $A$ contains no root then $m\geq 0$ and
 $\Delta$ 
 contains $m$ roots.
 \item If $m\neq 0$, then $A$ contains a root.
 \end{enumerate}
\end{Proposition}

We state the following corollary.

\begin{Corollary}[of Prop.~\ref{prop_CauchyRootCounterVerifHeu}]
\label{cor_CauchyRootCounterVerifHeu2}
 Let $\theta=4/3$ and $a=11/10$.
 Remark that
 \[
  f_-(a,\theta) = \frac{93}{110} > 2^{-1/4} \text{ and }
  f_+(a,\theta) = \frac{64}{55}.
 \]
 The call $\CauchyRootCounterVerifHeu(\OrP, \OrPp, \Delta, a)$ amounts to 
 $\lceil 2\pi\frac{a^2+1}{a^2-1}\rceil=67$
 calls to $\CauchyExclusion$,
 for discs of radius $\frac{21}{176}r\in O(r)$
 and one call to $\CauchyRootCounter$ for $\Delta$.
 This requires
 evaluation of $\OrP$ and $\OrPp$
 at
 $O(\cost{\deg, r})$
 points,
 and $O(\cost{\deg, r})$
 arithmetic operations, all 
 with precision less than $\prec{\deg, r}$.
\end{Corollary}

\section{Root radii algorithms}
 \label{sec_root_radii}
 
 \subsection{Approximation of the largest root radius}
\label{subsec_largest_root_radii}

For a monic $\pol$ of degree $\deg$ and 
bit-size $\tau=\log \| \pol \|_1$,
we describe a naive approach to the approximation
of the largest
modulus $r_\deg$ of a root of $\pol$.
Recall Cauchy's bound for such a polynomial: $r_\deg\leq 1+2^{\tau}$.
The procedure below finds
an $r$ so that $r_\deg<r$ and either
$r=1$ or $r/2 < r_\deg$
when $\pol$ is given by the evaluation oracles 
$\OrP, \OrPp$.

\begin{algorithmic}[1]
    \State $r\ass1$, $m\ass-1$
    \While {$m\leq \deg$}
        \State $m\ass \CauchyRootCounterVerifHeu(\OrP, \OrPp, \disc{0,r}, 4/3)$
        \If{$m<\deg$}
            \State $r\ass 2r$
        \EndIf
    \EndWhile
\end{algorithmic}
As a consequence of Prop.~\ref{prop_CauchyRootCounterVerifHeu}
each execution of the while loop terminates
and the procedure terminates after no more than $O(\tau)$ 
execution of the {\bf while} loop.
It requires evaluation of $\OrP$ and $\OrPp$
at $O(\tau \cost{\deg, r})$
points and 
$O(\tau \cost{\deg, r})$
arithmetic operations all with precision less than 
$\prec{\deg, r}$.
Its correctness is implied by correctness of the results of
$\CauchyRootCounterVerifHeu$
which is in
turn implied by correctness of the results of $\CauchyExclusion$.

\subsection{Approximation of the $(d+1-m)$-th root radius}
\label{subsec_dp1mm_root_radii}

For a $c\in\C$ and an integer $m\geq 1$, we call
$(d+1-m)$-th root radius from $c$ and write it
$r_{m}(c,\pol)$ the smallest radius of a disc centered
 in $c$ and containing exactly $m$ roots of $\pol$.

Algo.~\ref{algo:rootradius} approximates $r_{m}(c,\pol)$ within the relative error
$\nu$. It is based on the RC 
$\CauchyRootCounterVerifHeu$
and reduces the width of an initial interval $[l,u]$
containing $r_{m}(c,\pol)$ with a double exponential
sieve.

Its correctness for given input parameters is implied by correctness of the results of
$\CauchyRootCounterVerifHeu$
which is in
turn implied by correctness of the results of $\CauchyExclusion$.
Algo.~\ref{algo:rootradius} satisfies the proposition below\RO{, proved in \ref{app_subsec_proof_root_radii}}.

\begin{algorithm}[h!]
	\begin{algorithmic}[1]
	\caption{$\RootRadius(\OrP, \OrPp, \Delta, m, \nu,\varepsilon)$ }
	\label{algo:rootradius}
	\Require{ $\OrP$, $\OrPp$
	         evaluation oracles for $\pol$ and $\pol'$,
	         s.t. $\pol$ is monic of degree $\deg$.
	         A disc $\Delta=D(c,r)$, an integer $m\geq 1$,
	         $\nu\in\R$, $\nu>1$,
	         and $\varepsilon\in\R$ such that
	         $0<\varepsilon\leq r/2$
	         } 
	\Ensure{ $r'>0$ }
	\State choose $a$ s.t. 
	$\nu^{-\frac{1}{4}} < f_-(a,\thetaval) < f_+(a,\thetaval) < 2$
	\hfill {\it // when $\nu=2$ take $a=11/10$}
	
	\State $l\ass 0$, $u\ass r$
	
	\coblue{\it Find a lower bound to $r_{\deg+1-m}(c, \pol)$ }
	\State $m'\ass \CauchyRootCounterVerifHeu(\OrP, \OrPp, \disc{c,\varepsilon}, a)$
	\If{ $m'=m$}
        \State \Return $\varepsilon$
    \Else
        \State $l\ass f_-(a,\thetaval)\varepsilon$ 
    \EndIf
    
    \hspace{-1cm}\coblue{\it Apply double exponential sieve 
    to get $l \leq r_{\deg+1-m} \leq u \leq \nu l$ }
    \While{ $l < u/\nu$}
        \State $t\ass (lu)^{\frac{1}{2}}$
        \State $m'\ass \CauchyRootCounterVerifHeu(\OrP, \OrPp, \disc{c,t}, a)$
        \If{ $m'=m$ }
            \State $u\ass t$
        \Else
            \State $l\ass f_-(a,\thetaval)t$ 
        \EndIf
    \EndWhile
    \State \Return $u$
	\end{algorithmic}
\end{algorithm}
\vspace{-0.3cm}

\begin{Proposition}
 \label{prop_CauchyRootRadius}
 The call $\RootRadius(\OrP, \OrPp, \disc{c,r}, m, \nu, \varepsilon)$
 terminates after $O(\log \log(r/\varepsilon))$ iterations of the while loop.
 Let $\Delta=\disc{c,r}$ and $r'$ be the output of the latter call.
 \vspace{-0.2cm}
 \begin{enumerate}[(a)]
  \item If $\Delta$ contains at least a root of $\pol$ then so does $\disc{c,2r'}$.
  \item If $\Delta$ contains $m$ roots of $\pol$ and 
 $\CauchyRootCounterVerifHeu$ returns a correct result each time
 it is called in Algo.~\ref{algo:rootradius}, 
 then either $r'=\varepsilon$ and $r_m(c,\pol)\leq \varepsilon$,
 or 
 $r_m(c,\pol)\leq r' \leq \nu r_m(c,\pol)$.
 \end{enumerate}
\end{Proposition}
 
\section{A compression algorithm}
 \label{sec_compression}

\begin{figure}
	 \centering
	  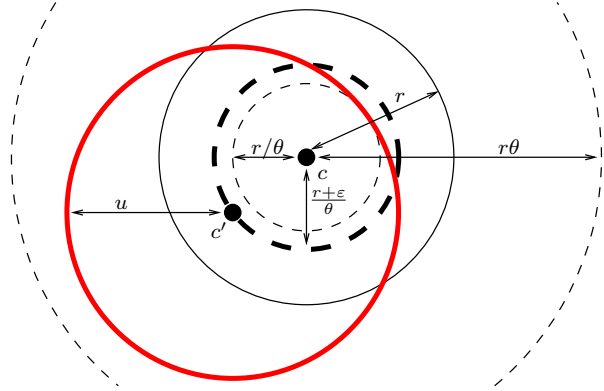
	    \caption{Illustration for Lem.~\ref{lem_eqdiscs} with
	    $\theta=2$ and $\varepsilon = r/4$.
	    $c'$ is on the boundary circle of $\disc{c,(r+\varepsilon)/2}$, and $u:=\left|c-c'\right| + r/\theta$.}
	 \label{fig:comp}
\vspace*{-0.3cm}
	\end{figure}

 We begin with a geometric lemma illustrated in Fig.~\ref{fig:comp}.
 \begin{Lemma}
 \label{lem_eqdiscs}
 Let $c\in\C$ and $r, \varepsilon, \theta \in\R$
 satisfying $0<\varepsilon \leq r/2$ and $\theta\geq 2$.
 Let $c'\in\disc{c, \frac{r+\varepsilon}{\theta}}$
 and $u=\max\left( \left|c-c'\right| + \frac{r}{\theta}, r\right)$.Then
 \[
 D\left(c,\frac{r}{\theta}\right) \subseteq 
 D\left(c', u\right)
 \subseteq
 D\left(c,\frac{7}{4}r\right)
 \subset
 D(c,r\theta).
\]
\end{Lemma}

\vspace*{-0.3cm}
	
The following lemma is a direct consequence of Lem.~\ref{lem_eqdiscs} because $\PowerSum{1}{\pol}{c}{r}/m$ is the center of gravity of the roots of $\pol$ in $\disc{c,r}$.

\begin{Lemma}
\label{lem_comp}
 Let $D(c,r)$ 
 be at least $\theta\geq 2$-isolated and contain $m$ roots.
 Let $s_1^*$ approximate $s_1(p, c, r)$ such that
 $|s_1^*-s_1(p,c,r)|\leq \frac{m\varepsilon}{\theta}$
 and 
 $\varepsilon \leq \frac{r}{2}$.
 Then for $c' = \frac{s_1^*}{m}$ and 
 $u=\max\left(\left|c-c'\right| + \frac{r}{\theta}, r \right)$,
 the disc $D(c',u)$ contains the same roots of $\pol$ as
 $D(c,r)$.  
\end{Lemma}

\vspace{-0.1cm}
Algo.~\ref{algo:compression} solves
the $\varepsilon$-CRD problem for $\gamma=1/8$.
It satisfies the proposition below
\RO{proved in \ref{app_subsec_proof_compression}}.

\begin{algorithm}[h!]
	\begin{algorithmic}[1]
	\caption{$\Compression(\OrP, \OrPp, \Delta, \varepsilon)$ }
	\label{algo:compression}
	\Require{ $\OrP$, $\OrPp$
	         evaluation oracles for $\pol$ and $\pol'$,
	         s.t. $\pol$ is monic of degree $\deg$. 
	         A disc $\Delta=\disc{c,r}$,
	         and a strictly positive $\varepsilon\in\R$.
	         } 
	\Ensure{ An integer $m$ and a disc $\disc{c',r'}$.}
	\State $\theta \ass 2$, $\varepsilon'\ass\varepsilon/2\theta$
	\State $(success, [\intbox{s_0}, \intbox{s_1}]) \ass \AppPhs(\OrP, \OrPp, \Delta, \theta, 1, \min(\varepsilon',1))$
	\If {{\bf not} $success$ {\bf or} $\intbox{s_0}$ does not contain an integer $>0$}
	 \State \Return $-1$, $\emptyset$
	\EndIf
	\State $m\ass$ the unique integer in $\intbox{s_0}$
	
	\If {$r/2 < \varepsilon$}
        \State \Return $m$, $\disc{c, r/2}$
    \EndIf
	\State $c'\ass \middle{\intbox{s_1}}/m$ 
	\hfill {\it // $|c' - s_1(\pol,c,r)/m|<\varepsilon/4\theta$}
	\If {$m=1$}
        \State $m\ass \CauchyRootCounter(\OrP, \OrPp, \disc{c', 2\varepsilon'}, 2)$
        \State \Return $m$, $\disc{c', 2\varepsilon'}$
	\EndIf
	\State $u\ass \max\left(|c-c'|+\frac{r}{\theta}, r\right)$
	\State $r'\ass \RootRadius(\OrP, \OrPp, \disc{c', u}, \frac{4}{3}, m, \theta, \varepsilon/2)$
    \State \Return $m$, $D(c', r')$
	\end{algorithmic}
\end{algorithm}

\begin{Proposition}
 \label{prop_compression}
 The call $\Compression(\OrP, \OrPp, \Delta, \varepsilon)$
 where $\Delta=\disc{c,r}$
 requires
 evaluation of $\OrP$ and $\OrPp$
 at 
 $O \left( \cost{\deg, \varepsilon}
  \logm{\logm{\frac{r}{\varepsilon}}}
    \right)$
 points and the same number of arithmetic operations, all
 with precision less than
 $\prec{\deg, \varepsilon/4}$.
%
 Let $m, \disc{c',r'}$ be the output of the latter call.
 \vspace{-0.2cm}
 \begin{enumerate}[(a)]
  \item If $\Delta$ is at least 2-isolated and $\solsIn{\Delta}{\pol}\neq \emptyset$,
 and if the call to $\RootRadius$ returns a correct result,
 then $\disc{c',r'}$ is equivalent to $\Delta$, contains $m$ roots of $\pol$
 and 
 satisfies: either $r'\leq \varepsilon$, 
 or $\disc{c',r'}$ is at least 
 $1/8$-rigid.
 \item If $m'>0$ then $\disc{c',2r'}$ contains at least a root of $\pol$.
 \end{enumerate}
%
\end{Proposition}

 
  \section{Two Cauchy Root Finders}
 \label{sec_root_finders}
 
 In order to demonstrate the efficiency of the algorithms presented
 in this paper,
 we describe 
 here two experimental subdivision algorithms,
 named $\cauchyN$ and $\cauchyC$, solving the $\varepsilon$-CRC
 problem for oracle polynomials 
 based on our Cauchy ET and RCs.
 Both algorithms can fail --in the case where 
 $\CauchyExclusion$ excludes a box of the subdivision tree containing a root -- but account for such a failure.
 Both algorithm adapt the subdivision process described 
 in \cite{becker2016complexity}.
 $\cauchyN$ uses QIR Abbott iterations to ensure fast convergence 
 towards clusters of roots.
 $\cauchyC$ uses $\varepsilon$-compression presented in 
 Sec.~\ref{sec_compression}.
 In both solvers, the main subdivision loop is followed
 by a post-processing step
 to check that the output is a solution
 of the $\varepsilon$-CRC problem.
 The main subdivision loop does not \forfinal{involve} coefficients of input polynomials
 but use evaluation oracles instead.
 However, we use coefficients obtained by evaluation-interpolation
 in the post-processing step in the case where some output discs
 contain more than one root.
 We observe no failure of our algorithms in all our experiments
 covered in Sec.~\ref{sec_experiments}.
 
 \subsection{Subdivision loop}
 \label{subsec_subd}
 
 Let $B_0$ be a box containing all the roots of $\pol$.
 Such a box can be obtained by applying the process described
 in Subsec.~\ref{subsec_largest_root_radii}.
 
 \vspace{-0.4cm}
 \subsubsection{Sub-boxes, component and quadrisection}
 \label{subsubsec_CC}
 
 For a box $B(a+\ii b,w)$, let 
 $Children_1(B)$ be the set of the four boxes
 $\{ B( (a\pm w/4) + \ii (b\pm w/4), w/2) \}$,
 and 
 \[
  Children_n(B) :=\bigcup_{B'\in Children_{n-1}(B)} Children_1(B').
 \]
 A box $B$ is a \emph{sub-box} of $B_0$
 if $B=B_0$ or if there exist an $n\geq 1$ s.t. $B\in Children_n(B_0)$.
 A \emph{component} $C$ is a set of connected sub-boxes of $B_0$
 of equal widths.
 The \emph{component box} $B(C)$ of a component 
 $C$ is the smallest (square) box subject to $C\subseteq B(C)\subseteq B_0$
 \RI{minimizing both $\Re(\middle{B(C)})$ and $\Im(\middle{B(C)})$}.
 We write $\contDisc{C}$ for $\contDisc{B(C)}$.
 If $S$ is a set of components (resp. discs) and $\delta>0$,
 write $\delta S$ for the set 
 $\{ \delta\contDisc{A} \text{ (resp. } A) ~|~ A\in S\}$.

 \begin{Definition}
 Let $Q$ be a set of components or discs.
 We say that a component $C$
 (resp. a disk $\Delta$)
 is $\gamma$-separated (or $\gamma$-sep.) from $Q$
 when $\gamma\contDisc{C}$
 (resp. $\gamma\Delta$)
 has empty intersection with all elements in $Q$.
\end{Definition}

\begin{Remark}
 Let $Q$ be a set of components and $C\notin Q$ a component.
 If $\solsIn{\C}{\pol}=\solsIn{\{C\}\cup Q}{\pol}$
 and $C$ is 4-separated from $Q$ then $2\contDisc{C}$ is at least $2$-isolated.
\end{Remark}
 
 \vspace{-0.5cm}
 \subsubsection{Subdivision process}
 \label{subsubsec_subdivision}
 
 We describe in Algo.~\ref{algo:cauchy} a subdivision
 algorithm solving the $\varepsilon$-CRC problem.
 The components in the working queue $Q$ are sorted 
 by decreasing radii of their containing discs.
 It is parameterized by the flag $compression$
 indicating whether compression or QIR Abbott iterations
 have to be used.
 In QIR Abbott iterations
 of Algo. 7 in \cite{becker2018near}, we replace
 the Graeffe Pellet test for counting roots in a disc $\Delta$ by 
 $\CauchyRootCounterVerifHeu(\OrP, \OrPp, \Delta, 4/3)$.
 If a QIR Abbott iteration in step 12 fails
 for input $\Delta, m$,
 it returns $\Delta$.
 Steps 20-21 prevent $C$ to artificially inflate when 
 a compression or a QIR Abbott iteration step
 does not decrease $\contDisc{C}$.
 For a component $C$,
 $Quadrisect(C)$ is the set of components
 obtained by grouping the set of boxes
 \[
  \bigcup_{B\in C} \left\{ B'\in Children_1(B) ~|~ 
  \CauchyExclusion(\OrP, \OrPp, \contDisc{B'}) = -1 \right\}
 \]
 into components.
 
 The {\bf while} loop in steps 4-22 terminates because
 all our algorithms terminate, and as a consequence of $(a)$
 in Prop.~\ref{prop_CauchyRootCounter}:
 any component will eventually be decreased until the 
 radius of its containing disc reaches $\varepsilon/2$.
 

 \begin{algorithm}[h!]
	\begin{algorithmic}[1]
	\caption{$\Cauchy(\OrP, \OrPp, \varepsilon, compression)$ }
	\label{algo:cauchy}
	\Require{ $\OrP$ and $\OrPp$
	         evaluation oracles for $\pol$ and $\pol'$,
	         s.t. $\pol$ is monic of degree $\deg$. 
	         A (strictly) positive
	         $\varepsilon\in\R$,
	         a flag $compression\in\{true, false\}$.
	         } 
	\Ensure{ A flag $success$ and a list $R=\{(\Delta^1,m^1),\ldots,(\Delta^\ell,m^\ell)\}$}
    \State $B_0\ass$ box s.t. $\nbSolsIn{B}{\pol}=\deg$
    as described in Subsec.~\ref{subsec_largest_root_radii}
    \State $Q\ass\{ B_0 \}$
    \hfill {\it // $Q$ is a queue of components
                   }
    \State $R\ass\{ \}$
    \hfill {\it // $R$ is the empty list of results}
    \While { $Q$ is not empty }
        \State $C\ass pop(Q)$
        \If { $C$ is 4-separated from $Q$}
            \If { $compression$}
                \State $m, \disc{c, r}\ass\Compression(\OrP, \OrPp, 2\contDisc{C}, \varepsilon/2)$
            \Else
                \State $m\ass \CauchyRootCounter(\OrP, \OrPp, 2\contDisc{C}, 2)$
                \If {$m>0$}
                    \State $\disc{c, r}\ass$ QIR Abbott iteration 
                    for $\contDisc{C},m$
                \EndIf
            \EndIf
            \If {$m\leq 0$}
                \State \Return $fail$, $\emptyset$
            \EndIf
            \If {$r\leq \varepsilon/2$ {\bf and} 
            $\disc{c, 2r}$ is $3$-sep. from $2Q$
            {\bf and} is $1$-sep. from $6Q$}
                \State $push(R, (\disc{c, 2r},m))$
                \State {\bf continue}
            \Else
                \State $C'\ass$ component containing $D(c,r)$
                \If {$C'\subset C$}
                    \State $C\ass C'$
                \EndIf
            \EndIf
        \EndIf
        \State $push(Q, Quadrisect(C))$
    \EndWhile
    \State $success\ass$ verify $R$ as described in \ref{subsec_output_verif}
    \State \Return $success, R$
	\end{algorithmic}
\end{algorithm}
 
 \vspace{-0.6cm}
 \subsection{Output verification}
 \label{subsec_output_verif}
 
 After the subdivision process described in 
 steps 1-22 of Algo.~\ref{algo:cauchy},
 $R$ is a set of pairs of the form
 $\{(\Delta^1,m^1),\ldots,\\(\Delta^\ell,m^\ell)\}$
 satisfying, for any $1\leq j\leq \ell$:
 \begin{itemize}
		 \item $\Delta^j$ is a disc of radii $\leq \varepsilon$, $m^j$ is an integer $\geq 1$,
		 \item $\Delta^j$ contains at least a root of $\pol$,
		 \item for any $1\leq j'\leq \ell$ s.t. $j'\neq j$,  
		 $3\Delta^j\cap\Delta^{j'}=\emptyset$.
 \end{itemize}
 The second property follows from $(b)$
 of Prop.~\ref{prop_CauchyExclusion}
 and $(b)$ of Prop.~\ref{prop_compression}
 when compression is used.
 Otherwise, remark that a disk $\Delta$ in the
 output of QIR Abbott iteration in step 12
 of Algo.~\ref{algo:cauchy} verifies
 $\CauchyRootCounterVerifHeu(\OrP, \OrPp, \Delta, 4/3)>0$
 and apply $(b)$ of Prop.~\ref{prop_CauchyRootCounterVerifHeu}.
 The third property follows from the {\bf if}
 statement in step 15 of Algo.~\ref{algo:cauchy}.
 Decompose $R$ as the disjoint union $R_1\cup R_{>1}$
 where $R_1$ is the subset of pairs $(\Delta^i,m^i)$ of $R$
 where $m^i=1$ and
 $R_{>1}$ is the subset of pairs $(\Delta^i,m^i)$ of $R$
 where $m^i>1$, and make the following remark:
 
 \begin{Remark}
 \label{rem_check_output}
 If $m^1+\ldots+m^\ell = \deg$ and for any $(\Delta^i,m^i)\in R_{>1}$, 
 $\Delta^i$ contains exactly $m^i$ roots
 of $\pol$, then $R$ is a correct output 
 for the $\varepsilon$-CRC problem with input $\pol$ of degree $\deg$ and $\varepsilon$.
\end{Remark}
 
 According to Rem.~\ref{rem_check_output},
 checking that $R$ is a correct output 
 for the $\varepsilon$-CRC problem for fixed input 
 $\pol$ of degree $\deg$ and $\varepsilon$
 amount to check that the $m^i$'s add up to $\deg$
 and that 
 for any $\Delta^i\in R_{>1}$,
 $\Delta^i$ contains exactly $m^i$ roots of $\pol$.
 For this last task, we use 
 evaluation-interpolation to approximate the coefficients
 of $\pol$ and then apply the Graeffe-Pellet test
 of \cite{becker2016complexity}.
 
 \section{Experiments}
 \label{sec_experiments}
 
 We implemented Algo.~\ref{algo:cauchy}
 in the \clang library \ccluster.
 Call \cauchyC (resp. \cauchyN) the implementation
 of Algo.~\ref{algo:cauchy} with $compression=true$
 (resp. $false$). In the experiments we conducted
 so far, \cauchyC and \cauchyN never failed.
 
 \vspace{-0.4cm}
 \subsubsection{Test suite}
 We experimented \cauchyC, \cauchyN and \mpsolve
 on Mandelbrot and Mignotte polynomials
 as defined in Sec.~\ref{sec_introduction}
 as well as Runnel and random sparse
 polynomials.
 Let $r=2$.
The Runnel polynomial is defined inductively as
\[
 \Run{0}(z)=1,~~\Run{1}(z)=z,~~
 \Run{k+1}(z) = \Run{k}(z)^r + z\Run{k-1}(z)^{r^2}
\]
It has real coefficients, a multiple root (zero),
and can be evaluated fast.
 We generate random sparse polynomials of degree $\deg$, bitsize $\tau$
and $\ell\geq 2$ non-zero terms as follows,
where $\pol_i$ stands for the coefficient of the monomial of degree $i$
in $\pol$:
$p_0$ and $p_\deg$ are randomly chosen in $[-2^{\tau-1},2^{\tau-1}]$,
then $\ell-2$ integers $i_1,\ldots,i_{\ell-1}$ are randomly chosen
in $[1,\deg-1]$ and $p_{i_1},\ldots,p_{i_{\ell-1}}$
are randomly chosen in $[-2^{\tau-1},2^{\tau-1}]$.
The other coefficients are set to $0$.

\vspace{-0.4cm}
\subsubsection{Results}

We report in tab.~\ref{table_timings_intro} results
of those experiments for Mandelbrot and Mignotte
polynomials with increasing degrees 
and increasing values of $\log_{10}(\varepsilon^{-1})$.
We account
for the running time $t$ for the three above-mentionned 
solvers. For \cauchyN (resp. \cauchyC),
we also give the number $n$ of exclusion tests in the subdivision
process, and the time $t_N$ (resp. $t_C$)
spent in QIR Abbott iterations (resp. compression).

Our compression algorithm allows smaller running times for 
low values of $\log_{10}(\varepsilon^{-1})$
because it compresses a component $C$ on the cluster it contains
as of $2\contDisc{C}$ is $2$-isolated, whereas QIR Abbott iterations
require the radius $\Delta$ to be near the radius of convergence
of the cluster for Schr\"oder's iterations.

We report in tab.~\ref{table_timings_final}
the results of runs of \cauchyC and \mpsolve for polynomials of our test suite
of increasing degree,
for $\log_{10}(\varepsilon^{-1})=16$.
For random sparse polynomials, we report averages over 10 examples.
The column $t_V$ accounts for the time spent in the verification
of the output of \cauchyC (see \ref{subsec_output_verif}); it is $0$ when all the pairs $(\Delta^j,m^j)$
in the output verify $m^j=1$.
It is $>0$ when there is at least a pair with $m^j>1$.

The maximum precision $L$
required in all our tests was 106, which makes us believe that 
our analysis in Prop.~\ref{prop_algo_compute_power_sums}
is very pessimistic.
Our experimental solver \cauchyC is faster than \mpsolve
for polynomials that can be evaluated fast.
 
 \begin{table}[h!]
\centering
 \begin{tabular}{c || c | c | c | c || c ||}
             & \multicolumn{4}{c||}{\cauchyC} 
             & \mpsolve \\\hline
$\deg$       & $t$  & $n$  & $t_C$ & $t_V$
             & $t$\\\hline 
\multicolumn{6}{c}{Mandelbrot polynomials} \\\hline
255 & \cored{1.31} & 5007 & 0.21 & 0.00 & \coblue{0.58} \\
511 & \coblue{3.25} & 10679 & 0.64 & 0.00 & \cored{4.13} \\
1023 & \coblue{6.47} & 18774 & 0.84 & 0.00 & \cored{31.7} \\
2047 & \coblue{16.2} & 39358 & 2.35 & 0.00 & \cored{267.} \\
\hline
\multicolumn{6}{c}{Runnels polynomials} \\\hline
341 & \cored{2.55} & 4967 & 0.38 & 0.00 & \coblue{0.45} \\
682 & \cored{5.66} & 9392 & 0.87 & 0.02 & \coblue{3.32} \\
1365 & \coblue{12.6} & 18030 & 2.00 & 0.05 & \cored{26.2} \\
2730 & \coblue{29.7} & 35612 & 4.26 & 0.12 & \cored{236.} \\
\hline
\multicolumn{6}{c}{Mignotte polynomials, $a=16$} \\\hline
256 & \cored{0.29} & 4131 & 0.15 & 0.00 & \coblue{0.21} \\
512 & \coblue{0.58} & 8042 & 0.27 & 0.00 & \cored{0.70} \\
1024 & \coblue{1.24} & 16105 & 0.55 & 0.02 & \cored{2.99} \\
2048 & \coblue{2.69} & 32147 & 1.05 & 0.04 & \cored{11.6} \\
\hline
\multicolumn{6}{c}{10 randomSparse polynomials with 3 terms and bitsize 256} \\\hline
767 & \cored{.902} & 10791. & .415 & 0.0 & \coblue{.602} \\
1024 & \coblue{1.35} & 15526. & .560 & 0.0 & \cored{1.36} \\
1535 & \coblue{2.04} & 21244. & .861 & 0.0 & \cored{2.35} \\
2048 & \coblue{2.98} & 30642. & 1.16 & 0.0 & \cored{4.10} \\
\hline
\multicolumn{6}{c}{10 randomSparse polynomials with 5 terms and bitsize 256} \\\hline
2048 & \cored{4.77} & 29583. & 1.60 & 0.0 & \coblue{4.09} \\
3071 & \coblue{6.92} & 43003. & 2.45 & 0.0 & \cored{10.0} \\
4096 & \coblue{9.82} & 56659. & 3.38 & 0.0 & \cored{24.0} \\
6143 & \coblue{17.7} & 86857. & 5.40 & 0.0 & \cored{44.5} \\
\hline
\multicolumn{6}{c}{10 randomSparse polynomials with 10 terms and bitsize 256} \\\hline
3071 & \cored{11.9} & 44714. & 4.09 & 0.0 & \coblue{10.3} \\
4096 & \coblue{17.5} & 58138. & 5.82 & 0.0 & \cored{17.6} \\
6143 & \coblue{29.1} & 85451. & 8.93 & 0.0 & \cored{51.9} \\
8192 & \coblue{40.6} & 116289. & 12.4 & 0.0 & \cored{66.5} \\
\hline
 \end{tabular}
 \caption{Runs of \cauchyC and \mpsolve on 
 polynomials of our test suite for $\log_{10}(\varepsilon^{-1})=16$.
 }
\label{table_timings_final}
\vspace*{-0.5cm}
\end{table}

 \section{Conclusion}
 \label{sec_conclusion}

 We presented, analyzed and verified practical
 efficiency of two basic subroutines for solving 
 the complex root clustering problem for black box 
 polynomials. One is a root counter, the other
 one is a compression algorithm. Both algorithms 
 are well-known tools used in subdivision procedures for root finding.
 
 We propose our compression algorithm not as a replacement
of QIR Abbott iterations, but rather as a complementary tool:
in future work, we plan to use compression to 
obtain a disc where Schr\"oder's/QIR Abbott iterations
would converge fast.

 
 The subroutines presented in this paper laid down the path
 toward a Cauchy Root Finder, that is, an algorithm solving the $\varepsilon$-CRC
 problem for black box polynomials.

\bibliographystyle{splncs04}
\bibliography{references}

\appendix

\section{Proofs}
\label{app_sec_proofs}

\subsection{Proof of Lem.~\ref{Lem_eval_bounds}}
\label{app_subsec_proof_eval_bounds}

Let $\Delta=\disc{c,r}$ contain $\deg_{\Delta}$ roots.
Suppose that the roots $\roo{1},\ldots,\roo{\deg}$ of $\pol$
are indexed such that 
$\roo{1},\ldots,\roo{\deg_{\Delta}}$ are in $\Delta$
and $\alpha_{d_{\Delta}+1},\ldots,\alpha_{d}$ are
outside $\Delta$.
Since $\Delta$ has isolation $\theta$, it follows that
\begin{align}
 \label{ineqa}
 |c+r\var^g-\roo{i}|&\geq r-\frac{r}{\theta} = \frac{r(\theta-1)}{\theta}
 \text{ when } i\leq \deg_{\Delta} \text{, and }\\
 \label{ineqb}
                       &\geq \theta r-r= r(\theta-1)
 \text{ when } i\geq \deg_{\Delta}+1
\end{align}
Write
\[
 \pol(c+r\var^g)=\lcf{\pol}\prod_{i=1}^{d_{\Delta}}(c+r\var^g-\alpha_i)
                      \prod_{i=d_{\Delta}+1}^{d}(c+r\var^g-\alpha_i)
\]
and deduce 
the first inequality of
Lem.~\ref{Lem_eval_bounds}.
Then write 
\[
 \frac{\pol'(c+r\var^g)}{\pol(c+r\var^g)}
 =\sum_{i=1}^{d_\Delta}\frac{1}{c+r\var^g - \alpha_i}
 +\sum_{i=d_{\Delta}+1}^{d}\frac{1}{c+r\var^g - \alpha_i}
\]
and deduce 
the second inequality of
Lem.~\ref{Lem_eval_bounds}.
\qed

\subsection{Proof of Lem.~\ref{lem_prec_approx_sh}}
\label{app_subsec_proof_lem_prec_approx_sh}

Lem.~\ref{lem_prec_approx_sh} is a direct consequence of the following 
Lemma:
\begin{Lemma}
\label{Lem_prec_approx_sh_2}
Let $\disc{c,r}$ be a complex disc,
$\theta>1$, $e>0$ and 
\[
 a = \max\left(\dfrac{\theta}{r(\theta-1)},1\right) ~~
\text{ and }
 \omega = \min\left(\dfrac{e}{26rda^{d+1}}, 1\right).
\]
Let $0\leq h < q$ be integers
and $\zeta$ be a primitive $q$-th root of unity.
For $0\leq g \leq q-1$, write 
$\psi^g = c+r\zeta^g$ and suppose that
\[
 |\pol(\psi^g)| \geq \frac{1}{2}\left(\frac{r(\theta-1)}{\theta} \right)^\deg
 \text{ and }
 \left|\frac{\pol'(\psi^g)}{\pol(\psi^g)}\right|
 \leq 2\left( \frac{\deg\theta}{r(\theta-1)}\right).
\]
For $0\leq g \leq q-1$ and $0\leq h < q$ write:
\[
 \zeta^{g(h+1)} = \app{\zeta^{g(h+1)}} + \delta_{\zeta^{g(h+1)}},
 \]
 \[
 \pol(\psi^g) = \app{\pol(\psi^g)} + \delta_{\pol(\psi^g)}
 \text{ and }
 \pol'(\psi^g) = \app{\pol'(\psi^g)} + \delta_{\pol'(\psi^g)}.
\]
Let
\[
 \app{s_h^*}=
                \frac{r}{q}\sum\limits_{g=0}^{q-1} 
                \app{\zeta^{g(h+1)}}
                \frac{\app{\pol'(\psi^g)}}
                {\app{\pol(\psi^g)}}
 \text{ and }
 s_h^* = \CauchySum{h}{q}{\pol}{c}{r}.
\]
If $|\delta_{\zeta^{g(h+1)}}|\leq \omega$,
   $|\delta_{\pol(\psi^g)}|\leq \omega$ and
   $|\delta_{\pol'(\psi^g)}|\leq \omega$
for all $0\leq g \leq q-1$ and $0\leq h < q$
then
\[
 |s_h^* - \app{s_h^*}|\leq e.
\]
\end{Lemma}

\noindent{\bf Proof of Lem~\ref{Lem_prec_approx_sh_2}}:
We depart from the inequality:
\begin{align*}
 \left| \frac{x}{y} -
  \frac{x+\delta_x}{y+\delta_y}
  \right|
  \leq
  \frac{|x\delta_y| + |y\delta_x|}{|y|(|y|-|\delta_y|)}
  & \leq
  \left( \left| \frac{x}{y} \right | +1 \right) \frac{\omega}{|y|-\omega} \\
  & \leq
  2 \left( \left| \frac{x}{y} \right | +1 \right) \frac{\omega}{|y|}
\end{align*}
valid for $|\delta_x|\leq \omega$,
and $|\delta_y|\leq \omega \leq \frac{1}{2}|y|$.
Substitute $x=\pol'(\psi^g)$ and $y=\pol(\psi^g)$
to obtain:
\begin{align*}
 \left| \frac{\pol'(\psi^g)}{\pol(\psi^g)}
 -\frac{\app{\pol'(\psi^g)}}{\app{\pol(\psi^g)}} \right|
  & \leq
 4(2\deg a + 1)a^{\deg}\omega\\
 & \leq 12\deg a^{\deg+1}\omega
 \text{ since } a\geq 1.
\end{align*}
Apply inequality
\begin{align*}
 | xy - (x+\delta_x)(y+\delta_y) | &\leq 
  | x\delta_y| + |y\delta_x| + |\delta_x\delta_y|
\end{align*}
to obtain
\begin{align*}
 \left| 
 \zeta^{g(h+1)}\frac{\pol'(\psi^g)}{\pol(\psi^g)}
 - \app{\zeta^{g(h+1)}}\frac{\app{\pol'(\psi^g)}}{\app{\pol(\psi^g)}} 
 \right|
 &\leq 12\deg a^{\deg+1}\omega
  + \left|\frac{\pol'(\psi^g)}{\pol(\psi^g)}\right|\omega
  + 12\deg a^{\deg+1}\omega^2\\
 &\leq 12\deg a^{\deg+1}\omega
  + 2\deg a\omega
  + 12\deg a^{\deg+1}\omega^2\\
 &\leq 26 \deg a^{\deg+1}\omega
 \text{ since } \omega\leq 1.
\end{align*}
and apply inequality
\[
 | x + y - ((x+\delta_x) + (y+\delta_y)) | \leq 
  |\delta_x| + |\delta_y|
\]
to get
\begin{equation*}
|s_h^* - \app{s_h^*}| 
\leq r26 \deg a^{\deg+1}\omega = e.
\end{equation*}
\qed
	
\subsection{Proof of Prop.~\ref{prop_algo_compute_power_sums}}
\label{app_subsec_proof_algo_compute_power_sums}

Suppose first that there is a $g$
so that $|\pol(c+r\zeta^g)|<\ell$.
The precision $L$ is increased by the while loop until
either $|\OrPEv{c+r\zeta^g}{L}|<\ell$
or $\frac{\ell}{2}\in|\OrPEv{c+r\zeta^g}{L}|$,
which holds for an $L\in O\left(\log 1/\ell\right) = O\left(\deg(\log\frac{\theta}{\theta-1} + \log \frac{1}{r})\right)$.

Similarly, suppose that there is a $g$
so that $\left|\frac{\pol'(c+r\zeta^{g})}
               {\pol(c+r\zeta^{g})}\right|>\ell'$.
The precision $L$ is increased by the while loop until
either 
$\left|\frac{\OrPpEv{c+r\zeta^{g}}{L}}
               {\OrPEv{c+r\zeta^{g}}{L}}\right|>\ell'$
or $2\ell'\in \left|\frac{\OrPpEv{c+r\zeta^{g}}{L}}
               {\OrPEv{c+r\zeta^{g}}{L}}\right|$,
which holds for an 
$L\in O\left(\log 1/\ell'\right) = O\left( \log\frac{r}{\deg} + \log\frac{\theta}{\theta-1}\right)$.

Thus Algo.~\ref{algo:approximating_phs} either terminates with 
$success=false$ for an 
\[L\in O\left( \max\left( \deg(\log\frac{\theta}{\theta-1} + \log \frac{1}{r}), \log\frac{r}{\deg} + \log\frac{\theta}{\theta-1} \right) \right)\]
or enters the for loop
with $|\OrPEv{c+r\zeta^g}{L}|>\frac{\ell}{2}$
and $\left|\frac{\OrPpEv{c+r\zeta^{g}}{L}}
               {\OrPEv{c+r\zeta^{g}}{L}}\right|
               <2\ell'$
for all $g$.
%
%
%
Then by Lem.~\ref{lem_prec_approx_sh},
Algo.~\ref{algo:approximating_phs}
terminates for $L\in O\left(
          \deg\left(\logm{\frac{1}{re}} + \log\frac{\theta}{\theta-1} \right)
    \right)$.


\begin{enumerate}[(a)]
 \item is a consequence of Lem.~\ref{Lem_eval_bounds}
 and eq. (\ref{eq_thm_app_psh_unit_disc})
 of Thm.~\ref{thm_app_ps_unit_disc}.
 \item If $\disc{c,r\theta}$ contains no root of $\pol$
 then (a) holds and for all $i\in\{0,\ldots, h\}$,
 $\PowerSum{i}{\pol_\Delta}{0}{1}=0$.
 \item If $\disc{c,r\theta}$ contains no root of $\pol$
 then (a) holds and $\PowerSum{0}{\pol}{c}{r}=\PowerSum{0}{\pol_{\Delta}}{0}{1}=\nbSolsIn{\Delta}{\pol}$.
 \item $success = false$ iff for a $g$
 it holds that 
 $|\OrPEv{c+r\zeta^g}{L}|<\ell$
or $\left|\frac{\OrPpEv{c+r\zeta^{g'}}{L}}
               {\OrPEv{c+r\zeta^{g'}}{L}}\right|>\ell'$
and so
$|\pol(c+r\zeta^g)|<\ell$
or $\left|\frac{\pol(c+r\zeta^{g'})}
               {\pol(c+r\zeta^{g'})}\right|>\ell'$
and $A(c,r/\theta, r\theta)$
contains at least a root of $p$.
 \item Suppose that $A(c,r/\theta, r\theta)$ contains no root of
 $\pol$ (otherwise the proposition is proved).
 Then $\PowerSum{i}{\pol_\Delta}{0}{1}\in\intbox{s_i}$
 and $\PowerSum{i}{\pol_\Delta}{0}{1}\neq 0$.
 Now if $\Delta$ contains no root of $\pol$ 
 then $\disc{0,1}$ contains no root of $\pol_\Delta$
 and $\PowerSum{i}{\pol_\Delta}{0}{1}=0$, which is a contradiction.
 \qed
\end{enumerate}

\subsection{Proof of Lemma~\ref{lem_circles}}
\label{app_subsec_proof_lem_circles}

See Fig.~\ref{fig:countingTest} for an illustration.
Let $c_{j,j+1}$ be the middle of $c_j$, $c_{j+1}$
and $z_j,z'_j$ be the intersections of the two circles
$C(c_j,a\rho), C(c_{j+1},a\rho)$ with $a > 1$,
such that $\|c-z_j\| < \|c-z_j'\|$.
Let $x$ be the distance $\|c_{j,j+1} - c_{j+1} \|$; one has
\begin{equation}
\label{eq:proof_Lem_circles2_eq1}
 x = \|c_{j,j+1} - c_{j+1} \| = \mu\sin(\pi/v) \leq \rho/2.
\end{equation}
Let $y$ be the distance $\|c_{j,j+1} - z_j \|$; one has
\begin{equation}
\label{eq:proof_Lem_circles2_eq2}
 y = \sqrt{ (a\rho)^2 - (\mu\sin(\pi/v))^2 } \geq 
     \rho \sqrt{ a^2 - 1/4}.
\end{equation}
where the inequality follows from Eq.~(\ref{eq:proof_Lem_circles2_eq1}).
Finally, 
\begin{equation}
\label{eq:proof_Lem_circles2_eq3}
\|c_{j,j+1} - c\| = \mu\cos(\pi/v) \geq
                   \mu\cos(1/2w). 
\end{equation}
One has to show:
\begin{align}
 \label{eq_goal1}
 \mu\cos(\pi/v) -y &\leq \mu - \rho \\
 \label{eq_goal2}
 \mu\cos(\pi/v) +y &\geq \mu + \rho
\end{align}
Eq.~(\ref{eq_goal1}) is straightforward when $y\geq\rho$,
which is the case when $a=5/4$.
According to inequalities~(\ref{eq:proof_Lem_circles2_eq2}) and~(\ref{eq:proof_Lem_circles2_eq3}),
Eq.~(\ref{eq_goal2}) holds if
\[
 \mu\cos(1/2w) + \rho \sqrt{ a^2 - 1/4}
 \geq \mu + \rho
\]
which rewrites
\[
 a \geq \sqrt{ (w(1-\cos(1/(2w)))+1)^2 + 1/4 }.
\]
The right-handside of the latter inequation is
decreasing with $w$ when $w\geq 1$, 
and is less that $5/4$ for any $w\geq 1$.
\qed


\subsection{Proof of Prop.~\ref{prop_CauchyRootRadius}}
\label{app_subsec_proof_root_radii}
Each call to $\CauchyRootCounterVerifHeu$ in Algo.~\ref{algo:rootradius}
terminates as a consequence of Prop.~\ref{prop_algo_compute_power_sums}.
Suppose that Algo.~\ref{algo:rootradius} enters the while loop with
\[
 l<\frac{u}{\nu}
 \text{ \emph{i.e.} }
 \frac{u}{l} > \nu.
\]
From the choice of $a$ in step 1, 
$f_-(a,\theta)>\nu^{-1/4}$ and one has
\[
 \frac{1}{f_-(a,\theta)}
 < \nu^{1/4} < \left(\frac{u}{l}\right)^{1/4}.
\]
Let $l',u'$ be the new values for $l,u$ after one iteration
of the {\bf while} loop;
in the worst case (step 14)
one has 
\[
  \frac{u'}{l'} 
 =  \frac{u}{ f_-(a,\theta) (lu)^{\frac{1}{2}} }
 = \left( \frac{u}{l} \right)^{1/2}
   \frac{1}{f_-(a,\theta)}
 < \left( \frac{u}{l} \right)^{1/2} \left(\frac{u}{l}\right)^{1/4}
 < \left( \frac{u}{l} \right)^{3/4}.
\]
As a consequence 
the {\bf while} loop
decreases
the value $\log_2\left(\frac{u}{l}\right)$
by at least $\frac{4}{3}$ every each recursive application
as long as 
\[
 \frac{u}{l}>\nu 
 \Leftrightarrow 
 \log_2 u - \log_2 l - \log_2 \nu > 0
\]
and so it stops in at most 
\[
 \lceil \log_{\frac{4}{3}} \left( 
           \log_2 u - \log_2 l - \log_2\nu         
        \right)
 \rceil
 \in O(\log \log(r/\varepsilon))
\]
steps.

Next, suppose that 
$\RootRadius(\OrP, \OrPp, \Delta, m, \nu, \varepsilon)$ return $r'$.

To prove $(b)$, suppose that
$\Delta$ contains $m$ roots of $\pol$ and 
$\CauchyRootCounterVerifHeu$ returns a correct result each time
it is called in Algo.~\ref{algo:rootradius}.

Suppose first that the call $\CauchyRootCounterVerifHeu$
in step 3 returns $m'=m$: then $r'=\varepsilon$ and 
$\disc{c,r'}$ contains $m$ roots thus 
$r_m(c,\pol)\leq \varepsilon$.

Suppose now that the call $\CauchyRootCounterVerifHeu$
in step 3 returns $m'\neq m$.
If $m\neq 0$, from Prop.~\ref{prop_CauchyRootCounterVerifHeu},
then $\annu{c,\varepsilon f_-(a,\thetaval), \varepsilon f_+(a,\thetaval)}$ contains a root.
Since $\varepsilon f_+(a,\thetaval) < 2\varepsilon \leq r$
(from the choice of $a$ in step 1)
and $\Delta=\disc{c,r}$ contains $m$ roots,
$r_m(c,\pol) >= f_-(a,\thetaval) \varepsilon$.

Finally, if the call $\CauchyRootCounterVerifHeu$
in step 3 returns $m'=0$ then 
$\disc{c,\varepsilon}$ contains no roots and 
$r_m(c,\pol) >= \varepsilon > f_-(a,\thetaval) \varepsilon$.

Thus if Algo.~\ref{algo:rootradius} enters the while loop,
it does so for $l$ and $u$ satisfying
$l<=r_m(c,\pol)<=u$.
Apply the same reasoning to show that
this property is preserved by the while loop,
and when it terminates,
$r_m(c,\pol)\leq r' \leq \nu r_m(c,\pol)$.

Let us finally prove $(a)$.
Suppose that $\Delta$ contains at least a root of $\pol$,
and let $r_1(c,\pol)$ be the smallest distance
of a root of $p$ to $c$.

Any call $\CauchyRootCounterVerifHeu(\OrP, \OrPp, \disc{c,t}, a)$
in Algo.~\ref{algo:rootradius}
with $t\leq (1/a)r_1(c,\pol)$ will 
necessarily return $0$ as a consequence
of Prop.~\ref{prop_CauchyRootCounterVerifHeu},
and $r'$ will necessarily be greater than
$(1/a)r_1(c,\pol)$.
Remark that the parameter $a$ chosen in step
1 is less than 2 and complete the proof.
\qed

\subsection{Proof of Prop.~\ref{prop_compression}}
\label{app_subsec_proof_compression}
Let $\theta=2$.
To prove $(a)$,
suppose that $\Delta$ is at least $\theta$-isolated.
Since $\min(\varepsilon',1)\leq 1$ one can apply
$(c)$ of Prop.~\ref{prop_algo_compute_power_sums}
and
after step 2 of Algo.~\ref{algo:compression},
$success$ is true and
$\intbox{s_0}$ contains the unique integer $m$
equal to the number of root of $\pol$ in $\Delta$.
As a consequence of $(a)$ of Prop.~\ref{prop_algo_compute_power_sums},
$\intbox{s_1}$ satisfies 
$|\middle{\intbox{s_1}} - s_1(\pol,c,r)|<\varepsilon'/2$
and since $m\geq1$,
$|\middle{\intbox{s_1}} - s_1(\pol,c,r)|<m\varepsilon/4\theta$.

If Algo.~\ref{algo:compression} enters
step 6, $\disc{c, r/2}$
contains $m$ roots and has radius less that $\varepsilon$.

Otherwise, $c'$ defined in step 8 satisfies 
$|c' - s_1(\pol,c,r)/m|<\varepsilon/4\theta$.
When $m=1$, the unique root of $\pol$ in $\Delta$
is $s_1(\pol,c,r)/m$,
thus $\disc{c', \varepsilon'}$
contains this root.

Suppose now $m\geq 2$; from Lem.~\ref{lem_comp},
the disc $\disc{c', u}$ 
where 
\[u=\max\left(\left|c-c'\right| + \frac{r}{\theta}, r \right)\] contains the same 
$m$ roots of $\pol$ as $\Delta$.
Also, $\varepsilon/\theta\leq u/2$ as required in Algo.~\ref{algo:rootradius}
for $\RootRadius$.
In step 13, $\RootRadius(\OrP, \OrPp, \disc{c', u}, \frac{4}{3}, m, \theta, \varepsilon/\theta)$
returns an $r'$ with either
$r'=\varepsilon/\theta$ and $r_m(c',\pol)\leq \varepsilon/\theta$,
 or 
 $r_m(c',\pol)\leq r' \leq \theta r_m(c',\pol)$.
 
 If $r'=\varepsilon/\theta$, then $D(c', r')$
 contains the same $m$ roots of $\pol$ as $\Delta$.
 
 Otherwise, $r'\geq \varepsilon/\theta$, we proove that $D(c', r')$ is at least $1/8$-rigid.
 
 $m\geq 2$, and
 $s_1(\pol,c,r)/m$ is the center of gravity of the $m$ roots of $\pol$ in 
 $\Delta$.
 The distance from $s_1(\pol,c,r)/m$ to any root of $\pol$ in $\Delta$
 is maximized when one root, say $\alpha$, has multiplicity $m-1$,
 and the other one, say $\alpha'$, has multiplicity $1$.
 In this case, 
 $|\frac{s_1(\pol,c,r)}{m} - \alpha'|\leq 
  |\frac{s_1(\pol,c,r)}{m} - \alpha |=\frac{m-1}{m}|\alpha-\alpha'|$.
  Generally speaking, one has
 \[
  r_m\left( \frac{s_1(\pol,c,r)}{m}, \pol \right) \leq 
  \frac{m-1}{m} \max_{\alpha, \alpha'\in\solsIn{\Delta}{\pol}}|\alpha-\alpha'|. 
 \]
 Since $\left| c' - \frac{s_1(\pol,c,r)}{m}\right| \leq \frac{\varepsilon}{4\theta}$,
 one has
 \[
  r_m\left( c', \pol \right) \leq 
  \frac{m-1}{m} \max_{\alpha, \alpha'\in\solsIn{\Delta}{\pol}}|\alpha-\alpha'|
  + \frac{\varepsilon}{4\theta},
 \]
 and since $r'\leq \theta r_m\left( c', \pol \right)$, one has
 \[
  r'\leq 
  \theta\frac{m-1}{m} \max_{\alpha, \alpha'\in\solsIn{\Delta}{\pol}}|\alpha-\alpha'|
  + \frac{\varepsilon}{4}.
 \]
 Since $r'>\varepsilon/2$, $r'-\varepsilon/4\geq r'/2$ and
 \[
  \max_{\alpha, \alpha'\in\solsIn{\Delta}{\pol}}\frac{|\alpha-\alpha'|}{2r'}
  \geq \frac{m}{(m-1)\times\theta\times4}
  \geq \frac{1}{8}.
 \]

 We now prove $(b)$: 
 suppose Algo.~\ref{algo:compression}
 enters step 7: if $m>0$, 
 $\disc{c,r}$ contains at least a root
 as a consequence of $(e)$ of Prop.~\ref{prop_algo_compute_power_sums}.
 Suppose Algo.~\ref{algo:compression}
 enters step 11:
 $\disc{c,r}$ contains at least a root
 as a consequence of $(e)$ of Prop.~\ref{prop_algo_compute_power_sums}.
 Otherwise, $(b)$ is a consequence of 
 $(a)$ of Prop.~\ref{prop_CauchyRootRadius}.
 
 We finally prove the computational cost.
 When $r/2\geq \varepsilon$, the call 
 \[\AppPhs(\OrP, \OrPp, \Delta, 2, 1, \min(\varepsilon/4,1))\] 
 in step 2
 requires evaluation of $\OrP$ and $\OrPp$
 at 
  $ O \left( \cost{\deg, r, \varepsilon, 2} \right) $
 points 
 and the same number of arithmetic operations, all
 with precision less than
  $\prec{\deg, r, \varepsilon, 2}$.
 In step 12, $u\in O(r)$ (see Lem.~\ref{lem_eqdiscs}), thus
 the call 
 \[
\RootRadius(\OrP, \OrPp, \disc{c', u}, \frac{4}{3}, m, \theta, \varepsilon/2)
\]
 amounts to 
 \[
  O \left( \logm{\logm{\frac{r}{\varepsilon}}}\right)
 \]
 calls
 $\CauchyRootCounterVerifHeu(\OrP, \OrPp, \disc{c,t}, a)$
 with $a=11/10$ and $t\geq \varepsilon/2$.
 From Cor.~\ref{cor_CauchyRootCounterVerifHeu2},
 this amounts to
 evaluation of $\OrP$ and $\OrPp$
 at 
 \[
  O \left( \cost{\deg, \varepsilon}
  \logm{\logm{\frac{r}{\varepsilon}}}
    \right)
 \]
 points and the same number of arithmetic operations, all
 with precision less than 
  $\prec{\deg,\varepsilon}$.
\qed

\end{document}

%% file: DetCountingGen3.pdf_t
\begin{picture}(0,0)%
\includegraphics{DetCountingGen3.pdf}%
\end{picture}%
\setlength{\unitlength}{4144sp}%
\begingroup\makeatletter\ifx\SetFigFont\undefined%
\gdef\SetFigFont#1#2#3#4#5{%
  \reset@font\fontsize{#1}{#2pt}%
  \fontfamily{#3}\fontseries{#4}\fontshape{#5}%
  \selectfont}%
\fi\endgroup%
\begin{picture}(3839,3728)(-3731,846)
\put(-855,3291){\makebox(0,0)[lb]{\smash{{\SetFigFont{8}{9.6}{\rmdefault}{\mddefault}{\updefault}{\color[rgb]{0,0,0}$c_1$}%
}}}}
\put(-877,2076){\makebox(0,0)[lb]{\smash{{\SetFigFont{8}{9.6}{\rmdefault}{\mddefault}{\updefault}{\color[rgb]{0,0,0}$c_{v-1}$}%
}}}}
\put(-719,2752){\makebox(0,0)[lb]{\smash{{\SetFigFont{8}{9.6}{\rmdefault}{\mddefault}{\updefault}{\color[rgb]{0,0,0}$c_0$}%
}}}}
\put(-2295,2752){\makebox(0,0)[lb]{\smash{{\SetFigFont{8}{9.6}{\rmdefault}{\mddefault}{\updefault}{\color[rgb]{0,0,0}$\rho_-$}%
}}}}
\put(-1934,2549){\makebox(0,0)[lb]{\smash{{\SetFigFont{8}{9.6}{\rmdefault}{\mddefault}{\updefault}{\color[rgb]{0,0,0}$c$}%
}}}}
\put(-3632,2764){\makebox(0,0)[lb]{\smash{{\SetFigFont{8}{9.6}{\rmdefault}{\mddefault}{\updefault}{\color[rgb]{0,0,0}$\rho = \frac{\rho_+ - \rho_-}{2}$}%
}}}}
\put(-3431,2090){\makebox(0,0)[lb]{\smash{{\SetFigFont{8}{9.6}{\rmdefault}{\mddefault}{\updefault}{\color[rgb]{0,0,0}$\rho_+$}%
}}}}
\put(-2548,2047){\makebox(0,0)[lb]{\smash{{\SetFigFont{8}{9.6}{\rmdefault}{\mddefault}{\updefault}{\color[rgb]{0,0,0}$\mu = \frac{\rho_+ + \rho_-}{2}$}%
}}}}
\put(-1573,3511){\makebox(0,0)[lb]{\smash{{\SetFigFont{8}{9.6}{\rmdefault}{\mddefault}{\updefault}{\color[rgb]{0,0,0}$\frac{5}{4}\rho$}%
}}}}
\put(-1277,2344){\makebox(0,0)[lb]{\smash{{\SetFigFont{8}{9.6}{\rmdefault}{\mddefault}{\updefault}{\color[rgb]{0,0,0}$\frac{2\pi}{v}$}%
}}}}
\end{picture}%

%% file: Comp.pdf_t
\begin{picture}(0,0)%
\includegraphics{Comp.pdf}%
\end{picture}%
\setlength{\unitlength}{4144sp}%
\begingroup\makeatletter\ifx\SetFigFont\undefined%
\gdef\SetFigFont#1#2#3#4#5{%
  \reset@font\fontsize{#1}{#2pt}%
  \fontfamily{#3}\fontseries{#4}\fontshape{#5}%
  \selectfont}%
\fi\endgroup%
\begin{picture}(3549,2311)(868,-4074)
\put(3790,-2676){\makebox(0,0)[lb]{\smash{{\SetFigFont{8}{9.6}{\rmdefault}{\mddefault}{\updefault}{\color[rgb]{0,0,0}$r\theta$}%
}}}}
\put(3172,-2367){\makebox(0,0)[lb]{\smash{{\SetFigFont{8}{9.6}{\rmdefault}{\mddefault}{\updefault}{\color[rgb]{0,0,0}$r$}%
}}}}
\put(2312,-2676){\makebox(0,0)[lb]{\smash{{\SetFigFont{8}{9.6}{\rmdefault}{\mddefault}{\updefault}{\color[rgb]{0,0,0}$r/\theta$}%
}}}}
\put(2709,-2830){\makebox(0,0)[lb]{\smash{{\SetFigFont{8}{9.6}{\rmdefault}{\mddefault}{\updefault}{\color[rgb]{0,0,0}$c$}%
}}}}
\put(2070,-3183){\makebox(0,0)[lb]{\smash{{\SetFigFont{8}{9.6}{\rmdefault}{\mddefault}{\updefault}{\color[rgb]{0,0,0}$c'$}%
}}}}
\put(1496,-3007){\makebox(0,0)[lb]{\smash{{\SetFigFont{8}{9.6}{\rmdefault}{\mddefault}{\updefault}{\color[rgb]{0,0,0}$u$}%
}}}}
\put(2656,-2986){\makebox(0,0)[lb]{\smash{{\SetFigFont{8}{9.6}{\rmdefault}{\mddefault}{\updefault}{\color[rgb]{0,0,0}$\frac{r+\varepsilon}{\theta}$}%
}}}}
\end{picture}%